\documentclass{amsart}
\usepackage{amssymb}
\newcommand{\ii}{\mathrm i}
\newcommand{\jj}{\mathrm j}
\newcommand{\kk}{\mathrm k}
\newcommand{\rep}{\mathbf}
\newcommand{\RR}{\mathbf R}
\newcommand{\CC}{\mathbf C}
\newcommand{\HH}{\mathbf H}

\title[The binary icosahedral group and grand unified theories]
{Possible uses of the binary icosahedral group in grand unified theories}
\author{Robert Arnott Wilson}
\date{First draft: 14th August 2021. This version: 2nd December 2021.}

\address{Queen Mary University of London}
\email{r.a.wilson@qmul.ac.uk}

\begin{document}
\begin{abstract}
There are exactly three finite subgroups of $SU(2)$ that act irreducibly in the spin $1$ representation, namely the binary tetrahedral,
binary octahedral and binary icosahedral groups.  In previous papers I have shown how the binary tetrahedral group gives rise to all
the necessary ingredients for a non-relativistic model of quantum mechanics and elementary particles, and how a modification of the
binary octahedral group extends this to the ingredients of a relativistic model. Here I investigate the possibility that the binary icosahedral group
might be related in a similar way to grand unified theories such as the Georgi--Glashow model, the Pati--Salam model,
various $E_8$ models and perhaps even M-theory.

This analysis suggests a possible way to combine the best parts of all these models into a new model that goes further than any of them
individually. The key point is to separate the Dirac spinor into two separate concepts, one of which is Lorentz-covariant, and the other
Lorentz-invariant. This produces a model which does not have a CPT-symmetry, and therefore supports an asymmetry between particles
and antiparticles. The Lagrangian has four terms, and there is a gauge group $GL(4,\RR)$ which permits the terms to be
separated in an arbitrary way, so that the model is generally covariant. Quantum gravity in this model turns out 
to be described by quite a different gauge group, namely $SO(5)$, acting on a $5$-dimensional space
that behaves like a spin $2$ representation after suitable symmetry-breaking.  
I then use this proposed model to make a few predictions and postdictions, and compare the results with experiment.
These suggest that the model is worth investigating further.
\end{abstract}
\maketitle

\tableofcontents
\section
{The group and its representations}
\subsection
{Introduction}
\label{intro}
In a series of papers \cite{perspective,finite,octahedral,gl23model} I have considered a number of options for using a finite group
in place of $SU(2)$ as the basis for a potential finite model of quantum mechanics. 
The basic idea is to study the group algebra, which is closely related to,  
but not identical to,
a Hilbert algebra with the group elements as a basis.
So far, the most promising option is the group
$GL(2,3)$ of all invertible $2\times 2$ matrices over the field of order $3$, that consists of the numbers $-1,0,1$ subject to the rule $1+1=-1$.
However, the quarks are not immediately evident in this model, so it is not necessarily the best of the available options.
There is just one further possibility that I have not explored in detail, namely the group $SL(2,5)$ of all $2\times 2$ matrices of determinant $1$
over the field of order $5$, that consists of the numbers $-2,-1,0,1,2$ with $1+2=-2$ and so on.

This group is also known as the binary icosahedral group $2I$, and is a double cover of the icosahedral group $I\cong Alt(5)$,
the alternating group on five letters.
Since  
it has a $5$-dimensional irreducible representation it maps into
$SU(5)$ and may therefore 
be related in some way to the Georgi--Glashow model \cite{GG}.
The group also has two unitary representations in two dimensions, and one in four dimensions, so that it embeds in the gauge group
\begin{align}
SU(2)_L\times SU(2)_R\times SU(4)
\end{align} 
of the Pati--Salam model \cite{PatiSalam}.
A further reason for thinking that this particular group might be useful is the fact that the `fermionic' part of the group algebra has a natural
structure as a $15$-dimensional space of quaternions. Each of these quaternions can be expressed in $3$ independent ways as a complex $2$-space,
which gives us a total of $45$ Weyl spinors, that is generally recognised as the exact number that are required for the 
elementary fermions in the standard model of particle physics. 

Of course, this triple-counting is not adopted in the standard model,
so it is not obvious that this process will actually work in practice. 
But it is worth a try, since this is the only one of the binary polyhedral groups
that 
has enough spinors to 
provide a realistic prospect of modelling quarks in full. 
If this does not work, then it more or less rules out the
possibility of a discrete model of quantum mechanics based on a finite analogue of $SU(2)$.

There has been some interest in using 
the group $2I$ in physics \cite{Dechant}, where its 
relationship
to $E_8$, also described in \cite{icosians},  
may be relevant for the development of potential $E_8$ models \cite{E8}.
There is however a widely held view that such a model is impossible,
since it does not appear to have enough spinors \cite{DG}. For further background on
this specific group see \cite{icosians,Klein,KY}. 
The first of these references will be referred to
frequently. 
For more general background on group theory and representation
theory see \cite{Zee,JamesLiebeck}. 

\label{reps}
\subsection{Classification of representations}
\label{basicreps}
The group $SL(2,5)$ 
has order $120$. It contains a scalar group of order $2$, and the quotient modulo scalars is isomorphic to the alternating
group $Alt(5)$ of order $60$. There are $9$ conjugacy classes of elements, represented by the following matrices:
\begin{align}
\pm\begin{pmatrix}1&0\cr 0&1\end{pmatrix},
\begin{pmatrix}0&1\cr-1&0\end{pmatrix},
\pm\begin{pmatrix}0&1\cr-1&-1\end{pmatrix},
\pm\begin{pmatrix}1&1\cr0&1\end{pmatrix},
\pm\begin{pmatrix}1&2\cr0&1\end{pmatrix}.
\end{align}
Writing these elements of the group as $\pm1$, $i$, $\pm w$, $\pm t$, $\pm s$ respectively, the character table \cite{Atlas} is as follows:
\begin{align}
\begin{array}{c|ccccc}
&\pm1 & i & \pm w & \pm t & \pm s\cr\hline
\rep 1&1&1&1&1&1\cr
\rep 3a&3&-1&0& \tau & \sigma\cr
\rep 3b&3&-1&0&\sigma&\tau\cr
\rep 4a&4&0&1&-1&-1\cr
\rep 5&5&1&-1&0&0\cr\hline
\rep 2a&\pm2 &0&\mp1 & \mp \tau &\mp \sigma\cr
\rep 2b&\pm 2 & 0 & \mp1 & \mp\sigma & \mp\tau\cr
\rep 4b&\pm 4 & 0 & \pm1 & \mp 1& \mp1\cr
\rep 6 & \pm 6 & 0 & 0 & \pm1 & \pm1\cr\hline
\end{array}
\end{align}
where $\tau=(1+\sqrt5)/2$ and $\sigma=(1-\sqrt5)/2$. The horizontal line separates the faithful (or `fermionic') 
representations (in which $-1$ acts as $-1$)
from the faithless (or `bosonic') 
representations (in which $-1$ acts as $+1$). 
\subsection{Faithful 
representations}
 All the 
faithful representations
are pseudo-real, which means they can be written as quaternionic matrices of half the size. In particular, there are two representations
as $1\times 1$ quaternion matrices, which are described in detail in \cite{icosians}. 
That source also gives some interesting $2\times 2$ quaternion matrices for $\rep 2a+\rep 2b$. In terms of the generators $i$ and $w$, we can take the
following:
\begin{align}
\rep2a:&\quad i\mapsto \ii, \quad w\mapsto (-1+\tau \ii +\sigma \jj)/2\cr
\rep2b:&\quad i\mapsto \ii, \quad w\mapsto (-1+\sigma \ii +\tau \jj)/2\cr
\rep2a+\rep2b:&\quad i\mapsto \begin{pmatrix} \ii &0\cr 0&\jj\end{pmatrix}, \quad w\mapsto \frac12 
\begin{pmatrix}-1+\ii&-1-\ii\cr 1-\ii &-1-\ii\end{pmatrix}
\end{align}
In this last case there is a set of five involutions that generate a group $Q_8\circ D_8$ normalized by $SL(2,5)$ and
denoted in \cite{icosians} by the following notation:
\begin{align}
\mathbf g = \begin{pmatrix}-1&0\cr 0&1\end{pmatrix},
\mathbf h = \begin{pmatrix}0&1\cr 1&0\end{pmatrix},
\mathbf i = \begin{pmatrix}0&-\ii\cr \ii&0\end{pmatrix},
\mathbf j = \begin{pmatrix}0&-\jj\cr \jj&0\end{pmatrix},
\mathbf k = \begin{pmatrix}0&-\kk\cr\kk&0\end{pmatrix}
\end{align}

By using the Pauli matrices in the standard way, the representations can be written as complex matrices as follows:
\begin{align}
\rep2a:&\quad i\mapsto \begin{pmatrix}\ii&0\cr 0&-\ii\end{pmatrix}, 
\quad w\mapsto \frac12\begin{pmatrix}-1+\tau \ii & \sigma\cr -\sigma & -1-\tau \ii\end{pmatrix}\cr
\rep2b:&\quad i\mapsto \begin{pmatrix}\ii&0\cr 0&-\ii\end{pmatrix}, 
\quad w\mapsto \frac12\begin{pmatrix}-1+\sigma \ii & \tau\cr -\tau & -1-\sigma \ii\end{pmatrix}\cr
\rep2a+\rep2b:&\quad i\mapsto\begin{pmatrix}\ii&0&0&0\cr 0&-\ii&0&0\cr0&0&0&1\cr 0&0&-1&0\end{pmatrix}, 
\quad
w\mapsto 
\frac{1+\ii}{2}\begin{pmatrix}\ii&0&-1&0\cr 0&-1&0&\ii\cr -\ii&0&-1&0\cr 0&1&0&\ii\end{pmatrix}
\end{align}
In addition, we have
\begin{align}
&\mathbf g \mapsto\begin{pmatrix}-1&0&0&0\cr 0&-1&0&0\cr 0&0&1&0\cr 0&0&0&1\end{pmatrix},
\mathbf h \mapsto\begin{pmatrix}0&0&1&0\cr 0&0&0&1\cr 1&0&0&0\cr 0&1&0&0\end{pmatrix},\cr
&\mathbf i \mapsto\begin{pmatrix}0&0&-\ii&0\cr0&0&0&\ii\cr \ii&0&0&0\cr 0&-\ii&0&0\end{pmatrix}, 
\mathbf j \mapsto\begin{pmatrix}0&0&0&-1\cr0&0&1&0\cr 0&1&0&0\cr -1&0&0&0\end{pmatrix}, 
\mathbf k \mapsto\begin{pmatrix}0&0&0&-\ii\cr0&0&-\ii&0\cr 0&\ii&0&0\cr \ii&0&0&0\end{pmatrix}.
\end{align}
For what it is worth, these matrices 
satisfy the same relations as the Dirac matrices $\gamma_5$, $\gamma_0$, $\ii\gamma_1$, $\ii\gamma_2$, $\ii\gamma_3$
respectively.  

The representation $\rep4b$ can also be written as $2\times 2$ quaternion matrices, as is done implicitly in \cite{Cohen}, case $O_1$,
and it is also possible to calculate some from the
information in \cite{icosians}, or directly from the definition
\begin{align}
\rep4b&=S^3(\rep2a)\cong S^3(\rep2b).
\end{align}
The following matrices are written with respect to a basis which seems to make this representation
about as clean and tidy as it is possible to get: 
\begin{align}
\rep4b:&\quad i\mapsto \begin{pmatrix}\ii&0\cr 0&\ii\end{pmatrix},\cr
&\quad w\mapsto  \frac16\begin{pmatrix}-3\ii-\sigma\jj+\tau\kk & 3 -\alpha\jj + \beta\kk\cr
3\ii-\beta\jj -\alpha\kk & 3+\tau\jj+\sigma\kk\end{pmatrix},
\end{align}
where $\alpha=1+\sigma$ and $\beta=3\tau-1$.
The group also contains the element
\begin{align}
u\mapsto \begin{pmatrix}(-1+\tau\jj+\sigma\kk)/2 &0\cr 0 & 1\end{pmatrix}
\end{align}
of order $3$, which might perhaps be useful for describing a weak interaction acting on a left-handed spinor only (see Section~\ref{subgroups}).
The group of diagonal matrices then has order $12$, and 
acts on the first coordinate as an irreducible subgroup of $SU(2)$, but on the second coordinate as a subgroup of $U(1)$. In particular,
there is no element corresponding to $u$ that acts on the `right-handed' part of the spinor.
Again, we may translate to $4\times 4$ complex matrices if we wish:
\begin{align}
\rep4b: &\quad i\mapsto \begin{pmatrix}\ii&0&0&0\cr 0&-\ii &0&0\cr 0&0&\ii & 0\cr 0&0&0&-\ii\end{pmatrix},\cr
&\quad w\mapsto \frac16\begin{pmatrix}-3\ii & -\sigma+\tau\ii &3&-\alpha+\beta\ii\cr
\sigma+\tau\ii & 3\ii &\alpha+\beta\ii&3\cr
3\ii&-\beta-\alpha\ii& 3& \tau+\sigma\ii\cr
\beta-\alpha\ii&-3\ii&-\tau+\sigma\ii& 3
\end{pmatrix}.
\end{align}
The final one of the faithful representations can be written as $3\times 3$ quaternion matrices in various ways, such as
\begin{align}
\rep6:&\quad i\mapsto\ii\begin{pmatrix}1&0&0\cr0&-1&0\cr0&0&-1\end{pmatrix}, \quad
w\mapsto\frac{-1+\tau \ii +\sigma \jj}{4}\begin{pmatrix}\sigma&-1&-\tau\cr -1&\tau&\sigma\cr \tau&-\sigma&-1\end{pmatrix}.
\end{align}

\subsection{Faithless representations}
The 
faithless representations are all real. 
The representations in dimension $3$ can be written as follows:
\begin{align}
\rep3a:&\quad i\mapsto\begin{pmatrix}1&0&0\cr0&-1&0\cr0&0&-1\end{pmatrix}, \quad
w\mapsto\frac12\begin{pmatrix}\tau&-1&-\sigma\cr -1&\sigma&\tau\cr \sigma&-\tau&-1\end{pmatrix}\cr\cr
\rep3b:&\quad i\mapsto\begin{pmatrix}1&0&0\cr0&-1&0\cr0&0&-1\end{pmatrix}, \quad
w\mapsto\frac12\begin{pmatrix}\sigma&-1&-\tau\cr -1&\tau&\sigma\cr \tau&-\sigma&-1\end{pmatrix}
\end{align}
Notice that these matrices are very similar to those given above for $\rep6$, which makes clear the following tensor product
structures:
\begin{align}
\rep6&= \rep2a\otimes \rep3b = \rep2b\otimes \rep3a.
\end{align}
The representation $\rep4a$ can be expressed as the quaternionic tensor product of $\rep2a$ and $\rep2b$.
If one just uses the complex tensor product, one obtains a complex representation which
obscures the underlying real structure.
If we take $\rep2a$ in the form of left-multiplications by the above quaternions, and $\rep2b$ as right-multiplications, then we obtain the
following matrices:
\begin{align}
\rep4a:&\quad i \mapsto \begin{pmatrix}1&0&0&0\cr0&1&0&0\cr 0&0&-1&0\cr0&0&0&-1\end{pmatrix},\quad
w\mapsto\frac14\begin{pmatrix}-1&\sqrt5&-\sqrt5&-\sqrt5\cr -\sqrt5&1&3&-1\cr \sqrt5&3&1&1\cr -\sqrt5 & 1 & -1 & 3\end{pmatrix}
\end{align}

The other 
faithless representation $\rep5$ is most easily written as a deleted permutation representation on $6$ points: 
\begin{align}
\rep1+\rep5:&\quad i\mapsto (1,2)(3,4),\quad w\mapsto (1,2,3)(4,5,6).
\end{align}
Another way to construct $\rep5$ is as a monomial representation,
which can be obtained from the permutation representation $\rep1+\rep4a$ 
by introducing some complex cube roots of unity, $\omega$ and $\bar\omega=\omega^2$, into the action:
\begin{align}
\rep1+\rep4a:&\quad i\mapsto (1,2)(3,4),\quad w\mapsto (1,3,5),\cr
\rep5:&\quad i\mapsto (1,2)(3,4),\quad w\mapsto (1,3,5)(2,\omega2,\bar\omega2)(4,\bar\omega4,\omega4).
\end{align}

In fact, the representations $\rep3a+\rep3b$ and  $\rep6$ can also be obtained as monomial representations,
by inserting scalars into the permutation representation $\rep1+\rep5$ so that
\begin{align}
\rep3a+\rep3b:&\quad i\mapsto (1,2)(3,4)(5,-5)(6,-6),\quad w\mapsto (1,2,3)(4,5,6).\cr
\rep6:&\quad i \mapsto (1,2,-1,-2)(3,4,-3,-4)(5,\ii5,-5,-\ii5)(6,-\ii6,-6,\ii6),\cr
&\quad w\mapsto (1,2,3)(4,5,6).
\end{align}
This version of $\rep6$ does not exhibit either of the tensor product structures. Indeed, it is a well-known hard problem in
computational representation theory to find a tensor product decomposition for a representation that is known to have one \cite{untensor}.

\section
{Subgroups and subalgebras}
\subsection{The structure of the group algebra}
\label{groupalgebra}
The 
faithless part of the group algebra is a direct sum of real matrix algebras, one for each irreducible representation:
\begin{align}
\RR + 2M_3(\RR) + M_4(\RR) + M_5(\RR).
\end{align}
The 
faithful part, on the other hand, is a direct sum of quaternion matrix algebras:
\begin{align}
2\HH + M_2(\HH) + M_3(\HH).
\end{align}
Since the gauge groups of the standard model and most Grand Unified Theories (GUTs) are compact, it is worth separating out the
compact subgroups of the group algebra into the faithless part
\begin{align}
SO(3)_L \times SO(3)_R \times SO(4) \times SO(5)
\end{align}
and 
the faithful part
\begin{align}
Sp(1)_L \times Sp(1)_R \times Sp(2) \times Sp(3),
\end{align}
noting the isomorphisms
\begin{align}
Sp(1)&\cong SU(2)\cong Spin(3),\cr
Sp(2)&\cong Spin(5).
\end{align}
Restricting from quaternions to complex numbers we therefore obtain
\begin{align}
U(1)\times U(1) \times U(2) \times U(3)
\end{align}
which after identification of the four scalar factors $U(1)$ becomes isomorphic to, but not necessarily equal to,
 the standard model gauge group.
 
Moreover, the finite group connects these various Lie groups
together, and therefore reduces the apparent symmetries.
In particular, the following well-known two-to-one maps are compatible with the finite group action:
\begin{align}
SU(2)_L &\rightarrow SO(3)_L,\cr
SU(2)_R&\rightarrow SO(3)_R,\cr
SU(2)_L\times SU(2)_R &\rightarrow SO(4),\cr
Sp(2)&\rightarrow SO(5).
\end{align}
In addition, there are embeddings
\begin{align}
SU(2)_L\times SO(3)_R &\rightarrow Sp(3),\cr
SU(2)_R \times SO(3)_L &\rightarrow Sp(3),
\end{align}
that are also compatible with the finite group action.

It is very important to realise, however, that these maps between compact groups do \emph{not} extend to maps
between the non-compact groups. In particular, we note that $Sp(2)$ embeds in $SL(2,\HH)$, whereas $SO(5)$ embeds
in $SL(5,\RR)$.  While $Sp(2)$ is just a double cover of $SO(5)$, there is no non-trivial homomorphism at all between
$SL(2,\HH)\cong Spin(5,1)$ and $SL(5,\RR)$. Of equal importance is the complex equivalent, restricting from $5$ dimensions to $3$,
where again there is no non-trivial homomorphism between $SL(2,\CC)\cong Spin(3,1)$ and $SL(3,\RR)$.
It is therefore of vital importance to distinguish correctly between the groups $SO(3,1)$ and $SL(3,\RR)$, and to interpret
them appropriately in both relativistic and quantum physics.

\subsection{Spinors and isospinors}
\label{spinors} 
First we need a $4$-dimensional faithful representation to hold the Dirac spinor. The two obvious choices are $\rep2a+\rep2b$ and $\rep4b$.
If we want an action of $Spin(3,1)\cong SL(2,\CC)$, then we have to use $\rep4b$. But if we want an invariant splitting into
left-handed and right-handed spinors then we have to use $\rep2a+\rep2b$. In order to implement the standard model, therefore, we may %will
need to consider both of them, and mix them together in some appropriate way.

The Dirac algebra must then appear as some mixture of the following versions of the square of the spinor:
\begin{align}
(\rep2a+\rep2b)\otimes(\rep2a+\rep2b)&= \rep1+\rep4a+\rep3a+\rep3b+\rep4a+\rep1,\cr
\rep4b\otimes \rep4b &=\rep1+\rep4a+\rep3a+\rep3b+\rep5,\cr
(\rep2a+\rep2b)\otimes\rep4b &= \rep5+\rep3a+\rep3b+\rep5.
\end{align}
The first looks to be the closest to the standard model, but does not have an action of the Lorentz group on the spinor.
The second does have an action of the Lorentz group, but incorporates some extra symmetries that are not in the standard model.
The last looks even stranger, as it contains no scalar representation at all. In both cases, however, the extra symmetries arise
from converting the permutation representation $\rep1+\rep4a$ into the monomial representation $\rep5$. This introduces a hidden
triplet symmetry, either once or twice.
One triplet symmetry might be useful for incorporating the colour symmetry of the strong force into the
Dirac algebra, while the other may be useful for extending the standard model to deal with three generations simultaneously.

The first two suggested analogues of the Dirac algebra split into symmetric and antisymmetric parts as follows:
\begin{align}
\Lambda^2(\rep2a+\rep2b) &= \rep1 + \rep1 +\rep4a,\cr
\Lambda^2(\rep4b) &= \rep1+\rep5,\cr
S^2(\rep2a+\rep2b) &= \rep3a+\rep3b+\rep4a,\cr
S^2(\rep4b) &= \rep3a+\rep3b+\rep4a.
\end{align}
Hence it is only in the antisymmetric part of the algebra that we see any difference between them.
This difference occurs in the top two degrees of the Dirac algebra, that is in the terms $\gamma_5$ and $\gamma_5\gamma_\mu$.
It therefore does not affect the implementation of quantum electrodynamics (QED), but only affects its 
unification with the weak interaction.

The third suggestion splits into left-handed and right-handed components as
\begin{align}
\rep2a\otimes \rep4b &=\rep3a+\rep5,\cr
\rep2b\otimes \rep4b &=\rep3b+\rep5.
\end{align}
Since this is an invariant splitting, it must be a splitting that defines the weak and/or strong force, so that $\rep2a$ and $\rep2b$ are technically isospin
representations rather than spin representations. Let us define $\rep2a$ to be left-handed and $\rep2b$ to be right-handed, as our notation
already suggests. To obtain the Dirac algebra as usually described we must
incorporate both $\rep2a\otimes \rep4b$ and $\rep4b\otimes \rep4b$, each of
which contains a copy of $\rep3a+\rep5$. 
Then these two copies of $\rep3a+\rep5$  
must be effectively identified with complex
multiples of each other
in order to implement the formalism of electroweak mixing.
At the same time, the symmetry of $\rep5$ must be broken to $1+4$ in some way. 

If we want a compact $SU(3)$ as a gauge group for the strong force, then the only available place for it is in $Sp(3)$,
but the action of $SU(3)$ necessarily breaks the symmetry of $SU(2)_L\times SO(3)_R$ and $SU(2)_R\times SO(3)_L$.
It therefore breaks the symmetry of the weak gauge group $SU(2)_L$, and of the generation symmetry group $SU(2)_R$,
if that is how we choose to interpret these groups. 
Both these examples of symmetry-breaking are features of the standard model,
which is a promising sign that this group algebra contains the features we need.
Indeed, there is no intrinsic relationship between the two tensor product structures, and therefore 
a wide range of possible interpretations.

\subsection{Vector and adjoint representations}
\label{adjoints}
The adjoint representations of symplectic groups arise from the
symmetric squares of the defining representations, so that they are associated to the representations
\begin{align}
S^2(\rep2a)&= \rep3a,\cr
S^2(\rep2b)&= \rep3b,\cr
S^2(\rep4b)&= \rep3a+\rep3b+\rep4a.
\end{align}
The `vector' representations on the other hand arise from anti-symmetric squares, in the forms
\begin{align}
\rep2a\otimes \rep2b &= \rep4a,\cr
\Lambda^2(\rep4b) &= \rep1+\rep5.
\end{align}
In particular, notice the breaking of symmetry from $SO(6)$ to $SO(5)$,
in the form of a restriction from 
\begin{align}
Spin(6)\cong SU(4)
\end{align}
 to $Spin(5)\cong Sp(2)$. 
This allows us 
to associate the adjoint representation of $SU(4)$ with the representation
\begin{align}
\rep3a+\rep3b+\rep4a+\rep5,
\end{align} 
which will play an important role
later on in this paper. But now we have a great deal more symmetry-breaking to
contend with.

In particular, there is a change of signature from compact $SU(4)$ to $SL(2,\HH)$, which generalises the Lorentz group
$SL(2,\CC)$. The compact part of this group is now represented in $\rep3a+\rep3b+\rep4a$, while the boosts lie in $\rep5$.
The Lorentz group therefore mixes either $\rep3a$ or $\rep3b$ with a $3$-dimensional subspace of $\rep5$. This can be achieved by
breaking the symmetry group from $SL(2,5)$ to $SL(2,3)$, for which all three of these $3$-dimensional representations are equivalent,
and can therefore be mixed in arbitrary proportions.

\subsection{Maximal subgroups}
\label{subgroups}
For the purposes of studying symmetry-breaking, we need to study subgroups of the binary icosahedral group.
There are three types of maximal subgroups, obtained from the icosahedron or its dual dodecahedron by fixing either
\begin{itemize}
\item one of the five cubes inscribed in the dodecahedron,
\item one of the six pairs of opposite faces of the dodecahedron, or vertices of the icosahedron,
\item one of the ten pairs of opposite vertices of the dodecahedron, or faces of the icosahedron.
\end{itemize}
In terms of the $2\times2$ matrices over the field of order $5$, we may define
\begin{align}
j:=\begin{pmatrix}-2&0\cr0&2\end{pmatrix},\quad
k:=\begin{pmatrix}0&2\cr 2&0\end{pmatrix}, \quad
v:=\begin{pmatrix}-2&-1\cr-2&1\end{pmatrix}.
\end{align}
Then $i,j,k,v$ generate a subgroup of the first type, $j,t$ one of the second type, and $k,w$ one of the third type.
In terms of quaternions, in the $\rep2a$ representation we can take
\begin{align}
j\mapsto \jj,\quad k\mapsto \kk,\quad v\mapsto (-1+\ii+\jj+\kk)/2, \quad t\mapsto (-\tau-\ii+\sigma\kk)/2.
\end{align}

The first subgroup is a copy of the binary tetrahedral group, $2T=2Alt(4)$ or $SL(2,3)$. The other two are binary dihedral groups,
sometimes called dicyclic groups, $2D_{10}$ and $2D_6=2Sym(3)$ respectively. When treating them individually, we do not
need to distinguish the $5$ copies of $2T$, the $6$ copies of $2D_{10}$ or the $10$ copies of $2D_6$. But when we come to
consider them together, it will matter which copy we take. The most important point is that a copy of $2T$ and a copy of $2D_6$ can
intersect in either a subgroup of order $6$ or a subgroup of order $4$. This fact turns out to be important for the properties
of electro-weak unification, and especially for the distinction between leptons and quarks.

All three of these maximal subgroups split the $\rep4b$ `Dirac spinor' representation into two halves, but the properties
of the two halves are quite different in the three cases. The character tables of the three groups are given by
\begin{align}
\begin{array}{cccc}
\pm 1 & i & \pm v & \pm v^2\cr\hline
1&1&1&1\cr
1&1&\omega&\bar\omega\cr
1&1&\bar\omega&\omega\cr
3&-1&0&0\cr\hline
\pm2&0&\mp1&\mp1\cr
\pm2&0&\mp\omega&\mp\bar\omega\cr
\pm2&0&\mp\bar\omega&\mp\omega\cr\hline
\end{array}\qquad
\begin{array}{cccc}
\pm1 &\pm t & \pm s & \pm j\cr\hline
1&1&1&1\cr
1&1&1&-1\cr
2&-\tau&-\sigma & 0\cr
2&-\sigma&-\tau & 0\cr\hline
\pm1 &\pm1 &\pm1 & \pm\ii\cr
\pm1 &\pm1 &\pm1 & \mp\ii\cr
\pm2 & \mp\tau & \mp\sigma & 0\cr
\pm2 & \mp\sigma & \mp\tau & 0\cr\hline
\end{array}\qquad
\begin{array}{ccc}
\pm 1 &\pm w & \pm k\cr\hline
1&1&1\cr
1&1&-1\cr
2&-1&0\cr\hline
\pm1 &\pm1 & \pm\ii\cr
\pm1 & \pm1 & \mp\ii\cr
\pm2 &\mp1 & 0\cr\hline
\end{array}
\end{align}
In the first two cases, the Dirac spinor splits as the sum of the last two $2$-dimensional representations: in the first case
these representations are complex conjugates of each other, analogous to the Weyl spinors in the standard model. 
In the second case, they are equivalent to the restrictions of $\rep2a$ and $\rep2b$.

But in the
third case the Dirac spinor splits as $1+1+2$, so that there is a splitting into $1+1$ and $2$ which cuts across the
complex conjugation symmetry. It seems likely, therefore, that this last case will be useful for explaining why and how the
left-handed and right-handed Weyl spinors behave so differently in the standard model. In terms of the quaternionic notation,
there are two significant copies of the complex numbers, defined by $\kk$ and $\tau\ii+\sigma\jj$, and it is the interplay between
these two that we need to investigate closely. In particular, $\kk$ defines a complex structure on the whole of the Dirac spinor,
while $\tau\ii+\sigma\jj$ defines a complex structure only on half the Dirac spinor. Unlike the standard model complex structures
defined by $i$ and $i(1-\gamma_5)$, however, these two complex structures anti-commute with each other. It is this anti-commutation
that allows the finite model to model three generations rather than just one.

\section{Modular representations}
\label{modular}
\subsection{Reduction modulo $3$}
\label{mod3}
The interplay between $\rep1+\rep4a$ and $\rep5$, as well as between $\rep2a+\rep2b$ and $\rep4b$, 
suggested by the above discussion can be studied in terms
of the $3$-modular representation theory, along the lines taken in \cite{gl23model}. This is because the differences appear only in the column
headed $w$, where the elements of order $3$ occur. The $3$-modular Brauer character table \cite{Alperin,ABC} is as follows:
\begin{align}
\begin{array}{c|ccccc}
&\pm1 & i &   \pm t & \pm s\cr\hline
1&1&1&1&1\cr
3a&3&-1& \tau & \sigma\cr
3b&3&-1&\sigma&\tau\cr
4a&4&0&-1&-1\cr\hline
2a&\pm2 &0 & \mp \tau &\mp \sigma\cr
2b&\pm 2 & 0 & \mp\sigma & \mp\tau\cr
6 & \pm 6 & 0  & \pm1 & \pm1\cr\hline
\end{array}
\end{align}
That is, we have just deleted the column headed $w$ and the rows $\rep5$ and $\rep4b$. 
I have also changed the font so as to distinguish between ordinary representations and modular representations.
There is an extra subtlety, though, in that in order to represent
$\tau$ and $\sigma$ we need a square root of $5$, or equivalently, modulo $3$, a square root of $-1$. Hence we need to work over the field
of order $9$ if we want to distinguish $3a$ from $3b$ and $2a$ from $2b$. On the other hand, if we want to work over the field of order $3$,
we can simply add these rows together, and amalgamate the last two columns, and work with the table
\begin{align}
\begin{array}{c|ccccc}
&\pm1 & i &   \pm t \cr\hline
1&1&1&1\cr
4c&4&0&-1\cr
6a&6&-2&1\cr
\hline
4d&\pm 4 & 0  & \mp1\cr
6b & \pm 6 & 0  & \pm1 \cr\hline
\end{array}
\end{align}

All the characters with degree divisible by $3$ split off into separate blocks.
The other projective indecomposable modules have structures 
\begin{align}
\begin{array}{c}
1\cr 4a\cr 1
\end{array}
\qquad
\begin{array}{c}
4a\cr 1\cr 4a
\end{array}
\qquad
\begin{array}{c}
2a\cr 2b\cr 2a
\end{array}
\qquad
\begin{array}{c}
2b\cr 2a\cr 2b
\end{array}
\end{align}
over the larger field,
and 
\begin{align}
\begin{array}{c}
1\cr 4c\cr 1
\end{array}
\qquad
\begin{array}{c}
4c\cr 1\cr 4c
\end{array}
\qquad
\begin{array}{c}
4d\cr 4d\cr 4d
\end{array}
\end{align}
over the prime field. Thus we have the same structure 
as in \cite{gl23model}, that is effectively three Weyl spinors
glued on top of each other. The only difference here is that there is a larger symmetry group to play with.

One curious feature of this larger symmetry group is that there is an embedding of $SL(2,5)$ in $SL(2,9)$,
where the field of order $9$ consists of the elements
\begin{align}
0,\pm1,\pm\ii, \pm1\pm\ii
\end{align}
satisfying the rules $1+1=-1$ and $\ii^2=-1$. Hence, over this field, the $2$-dimensional representations behave as though they were 
complex conjugates of each other.
In other words, at the discrete level we can treat $2a$ and $2b$ very much like Weyl spinors, swapped by the
field automorphism that negates $\ii$, as a finite analogue of complex conjugation. But this `complex conjugation'
only exists at the discrete level, and does not lift to complex conjugation at the continuous level.

Over the field of order $3$, the combined representation
$4d$ represents the finite analogue of a Weyl or Dirac spinor, and the constituents of its tensor square are
\begin{align}
4d\otimes 4d &\sim 1+4c+6a+4c+1.
\end{align}
The copy of $6a$ splits off as a direct summand, but it is not possible to split the scalars from the copies of $4c$.
Note that
\begin{align}
4c\otimes 4c &\sim 1+4c+6a+4c+1
\end{align}
also, so that for certain purposes we may not need to distinguish between $4c$ and $4d$. 
The constituents of some other representations are
\begin{align}
\Lambda^2(6a) &\sim1+4c+6a+4c,\cr
\Lambda^2(6b) &\sim  1+4c+1+4c+1+4c.
\end{align}

It may be that the series of groups
\begin{align}
SL(2,3) \subset SL(2,5) \subset SL(2,9)
\end{align}
has something interesting to tell us about the relationship between $SU(2)$ and $SL(2,\CC)$ in the standard model.
Here, $SL(2,3)$ is the finite (modulo $3$) analogue of $SU(2)$, and $SL(2,9)$ is the finite (modulo $3$)
analogue of $SL(2,\CC)$. In Lie group terms, there is nothing in between $SU(2)$ and $SL(2,\CC)$.
The curious fact that modulo $3$, and \emph{only} modulo $3$, there is another group in between, may have profound
consequences for the structure of the discrete symmetries of quantum mechanics.

\subsection{Reduction modulo $2$}
\label{mod2}
On reduction modulo $2$ we lose the distinction between fermionic and bosonic representations, and have just four irreducibles
\begin{align}
\begin{array}{cccc}
1&w&t&s\cr\hline
1&1&1&1\cr
2&-1&-\tau&-\sigma\cr
2&-1&-\sigma&-\tau\cr
4&1&1&-1\cr\hline
\end{array}
\end{align}
Again, over the field of order $2$ we have to combine $s$ and $t$, and $\tau$ and $\sigma$, to get
\begin{align}
\begin{array}{c|ccc}
&1&w&t\cr\hline
1&1&1&1\cr
4e&4&-2&-1\cr
4f&4&1&-1\cr\hline
\end{array}
\end{align}
Here, $4e$ is what remains of the Dirac spinor, and $4f$ is what remains of spacetime.
There is not much information left in this table, but for what it is worth, we have
\begin{align}
\Lambda^2(4e)&\sim 1+1+4f\cr
\Lambda^2(4f) &\sim 1+1+4e\cr
S^2(4e)&\sim 1+1+4e+4f\cr
S^2(4f)&\sim 1+1+4e+4f\cr
4e\otimes 4f &\sim  1+1+1+1+4e+4e+4e
\end{align}
In all cases, $4f$ splits off as a direct summand, but $4e$ and $1$ can glue together in quite complicated ways. In other words,
the mathematics may allow us to glue some scalars, such as mass and charge, onto the spinors, without affecting the spacetime representation.

\subsection{Reduction modulo $5$}
\label{mod5}
On reduction modulo $5$, the spacetime representation $\rep4a$ breaks up as $1+3$, and therefore blurs the distinction between
Minkowski spacetime and Euclidean spacetime. Similarly, $\rep6$ breaks up as $2+4$, and the distinction between left-handed $\rep2a$ (and $\rep3a$)
and right-handed $\rep2b$ (and $\rep3b$) disappears. The Brauer character table is as follows:
\begin{align}
\begin{array}{c|ccc}
&\pm 1 & i & \pm w\cr\hline
1&1&1&1\cr
3&3&-1&0\cr
5&5&1&-1\cr\hline
2&2&0&-1\cr
4&4&0&1\cr\hline
\end{array}
\end{align}
The representation $5$ is projective, so splits off into a block of its own.
The other four PIMs have the following structures:
\begin{align}
\begin{array}{c}
1\cr
3\cr
1
\end{array}\qquad
\begin{array}{c}
3\cr
1+3\cr
3
\end{array}\qquad
\begin{array}{c}
4\cr
2\cr
4
\end{array}\qquad
\begin{array}{c}
2\cr
2+4\cr
2
\end{array}
\end{align}

Possibly the most significant part of the reduction modulo $5$ is that it permits us to model a quantised version of
Lorentz transformations on either $1.3$ or $3.1$ in place of Euclidean spacetime in $\rep4a$.
Lifting back to the real representations, this corresponds to a `mixing' between $\rep4a$ and both $\rep1+\rep3a$ and $\rep1+\rep3b$.
We may then compare
\begin{align}
\Lambda^2(\rep4a)&=\rep3a+\rep3b\cr
\Lambda^2(\rep1+\rep3a)&=\rep3a+\rep3a\cr
\Lambda^2(\rep1+\rep3b)&=\rep3b+\rep3b
\end{align}
Here we see a subtle distinction between $\rep3a+\rep3b$ and $\rep3a+\rep3a$ (or $\rep3b+\rep3b$)
for the Euclidean and Minkowski versions of the electromagnetic field.

In the Minkowski case, both $\rep3a+\rep3a$ and $\rep3b+\rep3b$ 
can be regarded as complex $3$-spaces, on which 
the real quadratic form can be extended either to a complex quadratic form or to an Hermitian form.
In the former case, the group $SO(3,\CC)\cong SO(3,1)$ acts, together with a scalar $U(1)$,
 while in the latter case
the group is $U(3)$. In particular, we could use $\rep3a+\rep3a$ for electromagnetism and
Maxwell's equations,
and $\rep3b+\rep3b$ for the strong force. 

\section{Towards interpretation}
\subsection{Spacetime}
For the purposes of attempting to construct a physical model from the group algebra, 
we should not take too much notice of
the Lorentz group itself, and look instead at the bigger picture. 
But I do not want to interpret the extension of $Spin(3,1)$ to $Spin(5,1)$ as an extension of spacetime from $3+1$ dimensions
to $5+1$ dimensions, since that is unphysical. Instead I want to re-arrange the three $3$-dimensional representations in such a way that
spacetime can still be interpreted in $3+1$ dimensions. A Lorentzian signature is available in either $\rep1+\rep3a$ or $\rep1+\rep3b$,
and a Euclidean signature in both of these and also in $\rep4a$. On the other hand, the group algebra does not provide any
method for mixing a `time' coordinate in $\rep1$ with `space' coordinates in $\rep3a$ or $\rep3b$, so that spacetime is more naturally
associated with the representation $\rep4a$, and the corresponding algebra of $4\times 4$ real matrices, that contains $SL(4,\RR)$
and therefore $SO(3,1)$.
 
Classical physics and relativity do not mention spinors, so must be
expressible entirely in terms of the faithless part of the group algebra. General relativity \cite{thooft,GR1,GR2},  for example,
is constructed using rank $2$ tensors on a (Euclidean or) Minkowski spacetime. 
It is straightforward to calculate all the rank $2$ tensors for all the available combinations of spacetime representations,
as follows:
\begin{align}
\rep4a\otimes \rep4a &= \rep1+\rep4a+\rep3a+\rep3b+\rep5\cr
\rep4a\otimes (\rep1+\rep3a) &=\rep3b+\rep4a+\rep4a+\rep5\cr
\rep4a\otimes(\rep1+\rep3b) & = \rep3a+\rep4a+\rep4a+\rep5\cr
(\rep1+\rep3a)\otimes(\rep1+\rep3a) &=\rep1+\rep1+\rep3a+\rep3a+\rep3a+\rep5\cr
(\rep1+\rep3a)\otimes (\rep1+\rep3b) &= \rep1+\rep3a+\rep3b+\rep4a+\rep5\cr
(\rep1+\rep3b)\otimes(\rep1+\rep3b) &=\rep1+\rep1+\rep3b+\rep3b+\rep3b+\rep5
\end{align}
Both the first and the fifth of these are equivalent to the modified Dirac algebra $\rep4b\otimes \rep4b$, which suggests
that one or both of the representations $\rep4a\otimes\rep4a$ and
\begin{align}
(\rep1+\rep3a)\otimes (\rep1+\rep3b)
\end{align}
 may be able to provide some suggestions for how to implement
a quantisation of general relativity, albeit with some inevitable
changes to the interpretation.

\subsection{Labellings} 
By this stage we have identified  two possible different types of `spinors' in $\rep2a+\rep2b$ and $\rep4b$, and three 
possible different types of `vectors' in
$\rep1+\rep3a$, $\rep1+\rep3b$ and $\rep4a$.  The most likely interpretations are that $\rep4b$ contains Dirac spinors,
while $\rep2a+\rep2b$ contains some kind of `isospinors'. Similarly, $\rep4a$ is most likely to be interpreted as spacetime,
and/or its dual $4$-momentum, while $\rep1+\rep3a$ and $\rep1+\rep3b$ are more likely to be interpreted as 
force fields of some kind. Nevertheless, there may be other possibilities, and we should not pre-judge the issues.

All these characters have the property that the value on $i$ is $0$. Hence the tensor product
of any vector or spinor with any character that is $0$ on the elements of order $3$ and $5$ gives rise to a multiple of the
relevant half of the regular representation. It is straightforward to calculate that there are exactly two such characters,
namely
\begin{align}
\rep3a+\rep3b+\rep4a+\rep5&=\Lambda^2(\rep3a+\rep3b)\cr
&=\rep3a+\rep3b+\rep3a\otimes \rep3b,\cr
\rep1+\rep4a+\rep5+\rep5&=S^2(\rep5).
\end{align}

Either of these labellings can therefore be used to label the entire group algebra as spinors or vectors of various types
for various types
of particles.
The first option is the finite equivalent of  the adjoint representation of the gauge group $SL(2,\HH)$, so seems
suitable for a labelling of force mediators. This suggests using the other labelling for matter,
although we might prefer to follow the Georgi--Glashow labelling with the equivalent of $\rep5+\Lambda^2(\rep5)$,
which gives the
first labelling again.

The representation $\rep3a+\rep3b+\rep4a+\rep5$ is equivalent to the monomial representation of dimension $15$, and 
has many expressions in terms of smaller representations, which
may represent more primitive concepts:
\begin{align}
\label{bosonsplitting}
\rep3a+\rep3b+\rep4a+\rep5 &= \Lambda^2(\rep1+\rep5)=\Lambda^2(S^2(\rep3a))=\Lambda^2(S^2(\rep3b))=\Lambda^2(\Lambda^2(\rep4b))\cr
&=\rep5 + \Lambda^2(\rep5)\cr
&=\Lambda^2(\rep3a+\rep3b)=\Lambda^2(\Lambda^2(\rep4a))\cr
&= \rep3a\otimes \rep5= \rep3b\otimes \rep5\cr
&= \rep3a\otimes(\rep1+\rep4a)=\rep3b\otimes(\rep1+\rep4a).
\end{align}
The representation $\rep1+\rep4a+\rep5+\rep5$, equivalent to the permutation representation on $15$ letters,
also has a number of possible derivations: 
\begin{align}
\rep1+\rep4a+\rep5+\rep5&= S^2(\rep5)\cr
&= \Lambda^2(\rep6)=\Lambda^2(\rep2a\otimes \rep3b)=\Lambda^2(\rep2b\otimes \rep3a)\cr
&= \Lambda^2(\rep2a)\otimes S^2(\rep3b) + S^2(\rep2a)\otimes \Lambda^2(\rep3b)\cr
&= \rep1\otimes(\rep1+\rep5) + \rep3a\otimes \rep3b\cr
&= \Lambda^2(\rep2b)\otimes S^2(\rep3a) + S^2(\rep2b)\otimes \Lambda^2(\rep3a).
\end{align}

Note incidentally that by losing the Lorentz group we have lost the standard model distinction between fermions
and bosons. Hence in this more general case, some of the force mediators might be classified as fermions by the
standard model. Clearly, however, massive fermions cannot mediate forces, so this could only possibly apply to neutrinos.
Indeed, the distinction between the two halves of the group algebra appears rather to be a distinction between massive
particles in the faithful part, and massless particles in the faithless part. So let us use this as the defining property,
and sort out the fermion/boson distinction later.

\subsection{Labellings for vectors}
It seems obvious that we must identify $\rep3a+\rep3b$ with the electromagnetic field, and with the photon. Thus we have
a $3$-dimensional space for the momentum, in each of two helicities. The standard model identifies $\rep3a$ with $\rep3b$,
which splits $\rep4a+\rep5$ into $1+3+5$, hence identifying $3+5$ with gluons, and effectively throwing away the scalar.
The scalar could be thought of as a graviton, perhaps, but the description as $\rep3a\otimes \rep3b$ strongly suggests
that we should identify this space with three generations of neutrino momenta. In other words, it forces us to
decompose a gluon into a pair of neutrinos. 

If this is correct, then it implies that there is an invariant 
splitting of the neutrinos into two types, which might be interpreted as neutrinos and antineutrinos. However, in this discrete model
there are only half as many neutrino degrees of freedom as in the
standard model, so that the concepts of momentum, generation, and neutrino/antineutrino
are mixed together in a complicated fashion that depends on the individual observer.
This could explain the phenomenon of neutrino oscillation, at least at a conceptual level.

To see this in more detail, we have to tensor with the observer's spacetime, either in $\rep4a$ or $\rep1+\rep3a$ or $\rep1+\rep3b$.
Then we have
\begin{align}
\rep4a\otimes \rep4a &= \rep1+\rep3a+\rep3b+\rep4a+\rep5\cr
(\rep1+\rep3a)\otimes \rep4a&=\rep3b+\rep4a+\rep4a+\rep5\cr
(\rep1+\rep3b)\otimes \rep4a&=\rep3a+\rep4a+\rep4a+\rep5\cr
\rep4a\otimes \rep5 &=\rep3a+\rep3b+\rep4a+\rep5+\rep5\cr
(\rep1+\rep3a)\otimes \rep5 &=\rep3a+\rep3b+\rep4a+\rep5+\rep5\cr
(\rep1+\rep3b)\otimes \rep5&=\rep3a+\rep3b+\rep4a+\rep5+\rep5.
\end{align}
This seems to suggest that the particles in $\rep4a$ 
change the energy in some way, while those in $\rep5$ do not.
The former, then, are participating in the weak interaction to change the observer's measurement of the mass,
while the latter are participating in the strong force and/or gravity, with no change of mass.

The second and third equations show that the chirality of the weak interaction, that is the question of whether it is $\rep1+\rep3a$
or $\rep1+\rep3b$ that is mixed with $\rep4a$ on the right-hand side, may not be intrinsic, but may instead be 
determined by the observer's choice of 
$\rep1+\rep3a$ or $\rep1+\rep3b$ to parametrise spacetime on the left hand side. In other words, the chirality of the weak interaction
may be equal to the
chirality of the observer's motion relative to an inertial frame. While this conjecture is inconsistent with
the conventional understanding of the standard model, it is not inconsistent with experiment \cite{Wu},
since the experiment has only been done in one particular chirality of the motion of the experiment.

\subsection{Labellings for spinors} 
\label{fermions}
The total dimension of the 
faithful part of the real group algebra is $60$, but since 
it consists of quaternionic representations, we should really consider them as such, 
say with the quaternions acting by left-multiplication, and the finite group acting on the right. Then we have $15$ dimensions of
quaternions, splitting as an algebra into $1+1+4+9$, and as a representation into
\begin{align}
1+1+2+2+3+3+3.
\end{align} 
Each quaternion can represent a Weyl spinor, so that we have enough degrees of freedom for one generation of 
standard model fermions. 

In fact, of course, the standard model does not use the quaternion structure, instead choosing a particular identification of
$\HH$ with $\CC^2$. By varying the complex structure, therefore, we can in effect 
obtain three times as many degrees of freedom, and hence
have exactly the right number for a three-generation model.
I have suggested two possible labellings for two possible spinors, as follows:
\begin{align}
(\rep3ab+\rep4a+\rep5)\otimes \rep4b &= (\rep2a+\rep4b+\rep6) + (\rep2b+\rep4b+\rep6)\cr &\quad+(\rep4b+\rep6^2)+(\rep2ab+\rep4b+\rep6^2)\cr
(\rep3ab+\rep4a+\rep5)\otimes (\rep2a+\rep2b) &= (\rep2a+\rep4b)+\rep6+(\rep2b+\rep6)+(\rep4b+\rep6)\cr
&\quad + (\rep2b+\rep4b)+\rep6+(\rep2a+\rep6)+(\rep4b+\rep6)\cr
(\rep1+\rep4a+\rep5^2)\otimes \rep4b &=\rep4b + (\rep4b+\rep6^2) + (\rep2ab+\rep4b+\rep6^2)^2\cr
(\rep1+\rep4a+\rep5^2)\otimes (\rep2a+\rep2b) &= \rep2a + (\rep2b+\rep6) + (\rep4b+\rep6)^2\cr
&\quad + \rep2b + (\rep2a+\rep6) + (\rep4b+\rep6)^2.
\end{align}

In 
the labelling of $\rep2a$ with $\rep1+\rep4a+\rep5^2$,
we can possibly see $\rep2a$ as (massless) neutrinos, with $\rep6$ giving mass to $3$ generations of the
other particles, with electrons in the form
\begin{align}
\rep2b+\rep6&= \rep2b+\rep2a\otimes \rep3b\cr
&= \rep2b + \rep2b\otimes \rep3a\cr
&= \rep2b\otimes (\rep1+\rep3a)\cr
&= \rep2a \otimes \rep4a
\end{align}
so that we can take out a generation symmetry in $\rep3b$ and reinstate a Dirac spinor in $\rep2b+\rep2a$. 
Moreover, we see an intriguing relationship between $\rep4a$ acting on the left-handed Weyl spinor $\rep2a$, and $\rep1+\rep3a$ acting on
the right-handed Weyl spinor $\rep2b$. 
However, it seems more likely that, as already suggested, these representations are better interpreted as isospin representations.
Whichever way we look at it, this may 
have something 
interesting to tell us about the mixing of electromagnetism
with the weak force. It may also have something to say about the mixing with the strong force, and the three generations.

Similarly for the quarks, except that there now appears to be 
a `colour' symmetry in $\rep4b$ that is mixed up with $\rep2a$ or $\rep2b$. 
If this is a reasonable interpretation, then this structure enforces colour confinement via
\begin{align}
(\rep2a+\rep2b)\otimes \rep3b &= \rep6 + \rep2b+\rep4b
\end{align}
which has an extra copy of $\rep2b$ that can be used to define the charge.
Now we can see a potential new labelling of the 15 fermions, completely different from the standard model:
\begin{align}
\rep1&\leftrightarrow \nu\cr
\rep4a&\leftrightarrow e_R, e_L,\mu_L,\tau_L\cr
\rep5 &\leftrightarrow  u_R^r, u_R^g, u_L, c_L, t_L\cr
\rep5 &\leftrightarrow d_R^r, d_R^g, d_L, s_L, b_L.
\end{align} 
Only the left-handed spinors get generation labels, and only the right-handed spinors get colour labels,
with only $2$, rather than $3$, degrees of freedom.
Of course, this allocation may not be correct, 
but it gives a rough idea of how it might be possible to classify fermions in a scheme similar
to the Georgi--Glashow scheme, but with less profligacy of unobservable variables, so that all three generations can be incorporated
without increasing the size of the model.

It may be objected that the neutrinos have no generation label in this scheme. This is consistent with the experimental fact that the
generation of a neutrino depends on the observer \cite{oscillation,
neutrinos,SNO}. That is to say, in this model
the generation label on a neutrino comes from its interaction with
another particle, rather than from its intrinsic properties.

Another possibility is to use $\rep3a+\rep3b+\rep4a+\rep5$ as a labelling for spinors of type $\rep4b$. This gives a splitting of the $15$ quaternions into
$6+6+8+10$ complex numbers, which might be useful for modelling fermions in the form of $6$ leptons, $6$ quarks, $8$ spin $1/2$ baryons
(the baryon octet) and $10$ spin $3/2$ baryons (the baryon decuplet). 
However, this does not seem to match the spin $1/2$ representations $\rep2a$ and $\rep2b$ and the spin $3/2$ representation
$\rep4b$ consistently to the given particles. 

\section{Electro-weak aspects of the standard model}
\subsection{The Dirac equation}
\label{Diracequation}
In the proposed model, the Dirac equation is a tensor equation, which translates into the standard model as a matrix equation.
It must arise from a tensor product of a $4$-dimensional faithless representation, and a $4$-dimensional faithful representation,
followed by a projection back onto the latter. In the standard model it is the Clifford algebra that provides this projection, but here
it is the representation theory of the finite group that does this job. The options for the representations are 
$\rep4a$, $\rep1+\rep3a$ and $\rep1+\rep3b$
for the former, and $\rep4b$ and $\rep2a+\rep2b$ for the latter. These six cases 
all have a
projection of the required kind, 
but only in the first two cases is the projection uniquely defined.
\begin{align}
\rep4a\otimes \rep4b&=\rep4b+\rep6+\rep6\cr
\rep4a\otimes (\rep2a+\rep2b)&= (\rep2b+\rep2a)+\rep6+\rep6\cr
(\rep1+\rep3a)\otimes \rep4b&=\rep2a+\rep4b+\rep4b+\rep6\cr
(\rep1+\rep3b)\otimes \rep4b &=\rep2b+\rep4b+\rep4b+\rep6\cr
(\rep1+\rep3a)\otimes (\rep2a+\rep2b) &=(\rep2a+\rep2b)+(\rep2a+\rep4b)+\rep6\cr
(\rep1+\rep3b)\otimes (\rep2a+\rep2b)&= (\rep2a+\rep2b)+(\rep2b+\rep4b)+\rep6.
\end{align}

Hence the first two are the only cases
in which we could realistically hope to obtain a version of the Dirac equation, which confirms our
choice of $\rep4a$ as the spacetime representation.
Ignoring an overall scale factor in each case,
this gives us a gauge group
\begin{align}
SL(2,\HH)&=Spin(5,1),\cr
SL(1,\HH)\times SL(1,\HH)&=SU(2)\times SU(2),
\end{align}
respectively. Since the Dirac equation requires a Lorentz gauge in the form of the group $Spin(3,1)$, only the first
option is available to us. We therefore have to choose the Dirac $\gamma$ matrices inside $SL(2,\HH)$.
There is a wide range of options, but note that we do \emph{not} have a free choice within $SL(4,\CC)$, as might
be thought. 

In particular, we cannot use the complexification, and must make a careful choice between the $\gamma_\mu$ and
$\ii\gamma_\mu$ to ensure that we only have $15$ real dimensions, and not $16$ complex dimensions. These choices were
exhaustively studied in \cite{Clifford} for all possibilities for the signature of the gauge group. For our purposes,
with signature $(5,1)$, the most suitable option is $\gamma_0,\gamma_1,\gamma_2,\gamma_3,\ii\gamma_5$,
This requires us to modify the Dirac equation to use $\ii\gamma_5$ for the mass term, 
rather than the scalar $\ii$, but this change has essentially no effect on anything else.
A similar modification is proposed in \cite{Newman}.
 
 It is easiest to choose the four rotations as generators,
which means they must be anti-Hermitian, and the following is a suitable choice:
\begin{align}
\ii\gamma_5=\begin{pmatrix} 0&0&1&0\cr 0&0&0&1\cr -1&0&0&0\cr 0&-1&0&0\end{pmatrix},&\quad
\gamma_1=\begin{pmatrix}0&\ii&0&0\cr \ii&0&0&0\cr 0&0&0&-\ii\cr 0&0&-\ii&0\end{pmatrix},\cr
\gamma_2=\begin{pmatrix}0&-1&0&0\cr 1&0&0&0\cr 0&0&0&1\cr 0&0&-1&0\end{pmatrix},&\quad
\gamma_3=\begin{pmatrix}\ii&0&0&0\cr 0&-\ii&0&0\cr 0&0&-\ii&0\cr 0&0&0&\ii\end{pmatrix}.
\end{align}
(Note that the
 standard Bjorken--Drell convention \cite{Griffiths,BjorkenDrell} is not a suitable choice, since
 it does not respect the underlying quaternionic structure.)
It can then be easily checked that $\gamma_0$ is Hermitian, and therefore represents one of the boosts.
The other four boosts are $\ii\gamma_0\gamma_5$, $\gamma_0\gamma_1$, $\gamma_0\gamma_2$ and
$\gamma_0\gamma_3$, so that the gauge group mixes $\gamma_0$ with the other boosts into a vector
representation of $SO(5)$. All of this is consistent with the original proposal to modify the Dirac algebra to 
form the representation $\rep4b\otimes \rep4b$. The Lie algebra of $SO(5)$ maps onto $\rep3a+\rep3b+\rep4a$, extended by a
copy of the vector representation $\rep5$ to an image of the Lie algebra of $SO(5,1)$. 

This Lie algebra, moreover,
contains exactly the four `force' representations $\rep3a$, $\rep3b$, $\rep4a$ and $\rep5$, and might therefore be the Lie algebra
of the gauge group of a completely unified quantum theory of everything.
It contains $5$ energy terms, sufficient for kinetic energy and four types of potential energy
for the four forces. On the other hand, this algebra is neither
compact nor unitary, so this theory is not a Yang--Mills theory. The reason for this is that is contains the
`Lorentz gauge group' $SL_2(\CC)$, which is not compact. Moreover, it does not satisfy the
Coleman--Mandula theorem \cite{ColemanMandula}, and therefore cannot be a quantum field theory \cite{WoitQFT}
in the usual sense. Hence
the theory can \emph{only} make sense as a discrete theory, and \emph{cannot} be translated into the 
standard language of quantum field theory.

\subsection{Chiral Dirac matrices}
\label{chiralDirac}
As we saw in Section~\ref{Diracequation},
the Dirac matrices $\ii\gamma_5$, $\gamma_1$, $\gamma_2$, $\gamma_3$
represent a basis for the representation $\rep4a$, 
and they can be translated therefore into matrices acting on $\rep2a+\rep2b$ that swap the chiral components $\rep2a$ and $\rep2b$.
Alternatively, we can choose $\gamma_5$, $\ii\gamma_1$, $\ii\gamma_2$, $\ii\gamma_3$, which now represent boosts in
$SO(4,1)$, so must be represented by Hermitian matrices. It is reasonable to choose 
these as the matrices $\mathbf g$, $\mathbf i$,
$\mathbf j$, $\mathbf k$ given in Section~\ref{basicreps}.
But if we want to interpret $\ii\gamma_5$ as a mass term, it may be better to use $\mathbf h$ instead of $\mathbf g$
for this purpose. Mathematically, it makes no difference, so it is just a question of which is more convenient for
purposes of interpretation.
In this case, $\gamma_0$ acts as a scalar on each representation, and in order to preserve the chiral
components, we must restrict the Lie group generators to even products of the other matrices.

Without $\gamma_0$, the even products of the other Dirac matrices 
give us a copy of $Spin(4)$, such that $\gamma_0$ generates a copy of
$\RR$ that adjusts the scale factor between $\rep2a$ and $\rep2b$.
It can be readily checked that the splitting of $Spin(4)$ into $SU(2)_L\times SU(2)_R$ is defined by
the splitting:
\begin{align}
\ii\gamma_5\gamma_1+\gamma_2\gamma_3,
\ii\gamma_5\gamma_2+\gamma_3\gamma_1,
\ii\gamma_5\gamma_3+\gamma_1\gamma_2;\cr
\ii\gamma_5\gamma_1-\gamma_2\gamma_3,
\ii\gamma_5\gamma_2-\gamma_3\gamma_1,
\ii\gamma_5\gamma_3-\gamma_1\gamma_2.
\end{align}
This splitting therefore translates to a splitting defined by projections with $1\pm \gamma_0$, and therefore into a
splitting between positive and negative energy.
The standard model uses projections with $1\pm \gamma_5$ instead, but this leads to a difficult problem of interpretation,
since it is not obvious what these projections mean. They are clearly physically very different,
but 
it is not clear why. Swapping $\gamma_0$ and $\gamma_5$ in this way,
as the group algebra requires, separates positive and negative energy very clearly, and provides an
obvious reason why they behave completely differently.

It should also be noted that there is no action of the Lorentz group $SL(2,\CC)$ on the representation
$\rep2a+\rep2b$, although there is an action of $SO(3,1)$ on $\rep2a\otimes \rep2b=\rep4a$. 
The standard model does not distinguish between $SO(3,1)$ and $SL(2,\CC)$ here, and is therefore forced to
complexify the Dirac algebra in order to allow $SL(2,\CC)$ to act on the equivalent of $\rep2a+\rep2b$.
\label{chirality}

Physical chirality of spacetime arises in this model from the splitting of the anti-symmetric square of the spacetime representation $\rep4a$
into two chiral pieces as $\rep3a+\rep3b$. Hence the model directly and accurately models physical chirality on a Euclidean spacetime
that is fundamental to quantum reality. It does not necessarily model physical chirality on a Minkowski spacetime,
or on a macroscopic scale that is
appropriate to any particular observer. Nor does it directly model any change of chirality as between different observers, although
it does provide a gauge group $SL(4,\RR)$ with which to transform between different observers' choices of $SO(4)$.
Indeed, the interpretation of a fundamental discrete chirality in terms of a macroscopic continuous chirality runs into all 
the
usual problems, especially the 
measurement problem, discussion of 
which is beyond the scope of this paper.

\subsection{Electroweak mixing}
As I have already pointed out, although there are maps between the compact parts of the groups that act on $\rep2a+\rep2b+\rep4b$
and on $\rep3a+\rep3b+\rep4a(+\rep5)$, this map does not extend to the non-compact parts. 
There is no copy of $SO(3,\CC)\cong SO(3,1)$ in the group
algebra, that acts on $\rep3a+\rep3b$. But there is a copy of 
\begin{align}
SO(4)/Z_2 \cong SO(3)_L\times SO(3)_R.
\end{align}
Hence, by complexifying the Dirac algebra one can identify $SO(3,1)$ with $SO(4)$, factor out the central involution from both groups,
and compute the correct answers with the resulting formalism.
This is in effect what the standard model does, and this formalism is indeed capable of producing the same answers
to the calculations that the group algebra model produces. 

But in the group algebra model, the adjoint representation $\rep3a+\rep3b$ of $SO(4)$ is different from the adjoint representation
of $SO(3,1)$, which has its compact part in $\rep3a+\rep3b$ and its boosts in $\rep4a+\rep5$. In particular, there is a mixing angle
between adjoint $SU(2)_L$ in $\rep3a$ and spin $SU(2)$ in $\rep3a+\rep3b$, known as the electroweak mixing angle.
This angle clearly depends on the energy, which changes the imposed complex structure on the rest of $\rep3a+\rep4a$.
Nevertheless, with the exception of the running of the mixing angle, the standard model does an excellent job of
describing what happens in electroweak interactions, as studied in the laboratory. 

Now extending from spin $SU(2)$ to $SL(2,\CC)=Spin(3,1)$ acting on $\rep4b$, 
we extend from $Spin(4)$ to $Spin(4,1)$, whose adjoint representation consists of $\rep3a+\rep3b$ together with a 4-space inside $\rep4a+\rep5$.
Therefore we also have a mixing between $\rep4a$ and $\rep5$ defined by the embedding of our choice of $SO(4,1)$ in $SO(5,1)$.
The only reasonable interpretation of this symmetry-breaking is that $SO(4,1)$ defines our concept of charge, regarded as fixed,
while allowing mass
to vary as it does in the weak interaction, so that $SO(3,1)$ then defines our concept of mass.

For the purposes of electroweak unification we only need $SO(4,1)$, with a fixed definition of charge, and a variable mass.
Then we have an action of
$SO(4,1)$ as symmetries of the Dirac equation, as a quantum version of
Einstein's equation
\begin{align}
m^2c^4+p^2c^2-E^2&=0.
\end{align}
But the finite group treats the energy as a scalar, and therefore strictly enforces not only the principle of
conservation of energy, but also the principle of invariance of energy. In other words, the model proposes that there
should be a definition of energy that does not depend on the observer. But if energy does not depend on the observer,
then mass must depend on the observer. 

Since the gauge group is $SO(4,1)$, this is not actually a physical distinction
at all, it is purely a choice of gauge. One can choose a Lorentz gauge, or one can choose an $SO(4)$ gauge.
For a finite group model, we have to choose a compact gauge group, namely $SO(4)$.
This contains both $SU(2)_L$ and $SU(2)_R$ as subgroups, the former implementing the change of mass seen
experimentally in
weak interactions.

Moreover, we can now see roughly
how the values of mass that are used in the standard theory
may depend, at least in some instances, on the observer's choice of Lorentz gauge. 
They depend on the electroweak mixing between $\rep3a$ and $\rep3b$,
that is between 
left-handed and right-handed angular momentum. Since the group algebra model is independent
of scale, this mixing scales up to astronomical scales. Masses of elementary particles are then related to the mixing of 
angular momentum on a Solar System scale. In other words, they are related to rotation, acceleration and the gravitational field.
But I emphasise again, that the model does \emph{not} predict that particle masses change \emph{physically} 
when the gravitational field changes. All that changes is the appropriateness or otherwise of particular gauges
in particular circumstances.

\section{Other aspects of the standard model}
\subsection{Helicity and chirality}
\label{chiralities}
Chirality in the standard model is expressed by splitting the Dirac spinor representation $\rep4b$ into two Weyl spinors.
Any such splitting in the finite model can only be achieved by breaking the symmetry, and restricting to a subgroup.
As we saw in Section~\ref{subgroups}, there are three types of maximal subgroups, which split the Dirac spinor into two in
three distinct ways. There are therefore three distinct concept of `chirality', which we must match up to the standard model
concepts of helicity and chirality.

First, there is the restriction to the subgroup $SL(2,3)$ of index $5$, that is the binary tetrahedral group. In this case, the Dirac spinor
restricts to a sum of two complex $2$-spaces. As representations, these are complex conjugates of each other, so that there is a
natural interpretation as left-handed and right-handed Weyl spinors, acted on by the subgroup $SL(2,\CC)$ of $SL(2,\HH)$.
The complex structure here is defined in the character of the representation by a cube root of unity, for example
$(-1+\ii+\jj+\kk)/2$, and an imaginary quaternion perpendicular to it, say $(\ii-\sigma\jj-\tau\kk)/2$, that effects complex conjugation.
However, it is not possible to write the finite group representations without introducing some additional irrationalities, such as $\sqrt3$.
This type of chirality corresponds to the chirality of $SL(2,\CC)$ in the standard model, which we might call Minkowski chirality or M-chirality.
The complex scalar multiplication group $U(1)$ that acts as the gauge group of electromagnetism in the standard model
is the centraliser of $SL(2,\CC)$ in $SL(2,\HH)$, but does not correspond to any subgroup of the finite group.

Second, there is the restriction to the subgroup of index $6$, that is a dicyclic group of order $20$, generated for example by $j$ and $t$.
In this case, the Dirac spinor restricts to a sum of two quaternionic $1$-spaces, which are equivalent to the restrictions of $\rep2a$
and $\rep2b$. These are acted on by a copy of $Spin(4)$, so that this chirality is a Euclidean chirality, or E-chirality.
In the standard model, the complex structure of the Dirac algebra
ensures that M-chirality and E-chirality are identified with each other, 
or combined into a single concept of EM-chirality. This might be thought of as `electromagnetic chirality',
and is usually called helicity.

Third, there is the restriction to the subgroup of index $10$, that is a dicyclic group of order $12$, generated for example by $k$ and $w$.
In this case the Dirac spinor restricts to a sum of two complex $1$-spaces and a quaternionic $1$-space. The ambient Lie group here
is therefore $U(1)\times SU(2)$, so that this type of chirality might be called weak chirality, or W-chirality. In the action of the finite group,
$k$ maps into both $U(1)$ and $SU(2)$, so creates a `mixing' between $U(1)$ and $SU(2)$. Note, however, that $k$ itself cannot be 
directly identified with a generator for $U(1)_{em}$, since $k$ acts by right-multiplication and $U(1)_{em}$ acts by left-multiplication.

The word chirality is usually
%then 
reserved for the concept of W-chirality. The finite model makes the distinctions rather more clearly than the standard model does, 
and associates each of the three
types of chirality with a particular type of symmetry-breaking. In so doing, it associates each of the forces with a particular subgroup
(up to conjugacy)
of the binary icosahedral group. Thus electromagnetism is associated with $SL(2,\CC)$ and therefore with the binary tetrahedral group,
the weak force with $U(1)\times SU(2)$ and therefore with the dicyclic group of order $12$, leaving 
$Spin(4)$
and the dicyclic group of order $20$ for the strong force and/or quantum gravity.

As we have seen, the representation $\rep4b$ is most naturally written with respect to a basis that exhibits the W-chirality. The standard
model of electroweak unification attempts to write W-chirality and M-chirality with respect to the same basis, which is essentially
impossible. The mathematical structure of this representation does not lend itself to a nice description of either M-chirality or E-chirality.

\subsection{The strong force}
To obtain the remaining forces, we have to extend from the $SO(4,1)$ gauge group to $SO(5,1)$. This adds a copy 
of the representation $\rep5$ to what is already there. Moreover, it provides a 
compact gauge group $SO(5)$ acting on $\rep5$ in its
vector representation, and on $\rep3a+\rep3b+\rep4a$ in its adjoint representation.
In other words, we can interpret $\rep5$ as a set of five `colours' for the strong force, or we can extend to six colours in $\rep1+\rep5$,
with the finite group acting as permutations of the colours, and therefore enforcing colour confinement by throwing away the scalar
(energy) component.

What the standard model does instead is to take $3$ colours and $3$ anticolours, thereby replacing $\rep1+\rep5$ either by 
$\rep3a+\rep3b$
or $\rep 6$. If the gauge group for the strong force is $SU(3)$, we must have $\rep6$. For compatibility with the group algebra model, 
we have to compare:
\begin{align}
\Lambda^2(\rep1+\rep5)&=\rep3a+\rep3b+\rep4a+\rep5,\cr
\Lambda^2(\rep3a+\rep3b)&=\rep3a+\rep3b+\rep4a+\rep5,\cr
\Lambda^2(\rep6)&=\rep1+\rep4a+\rep5+\rep5.
\end{align}
From this we see that any choice of $SU(3)$ breaks the symmetry again. Such symmetry-breaking cannot be obtained
by restricting to a subgroup of the finite group, since no proper subgroup has a 3-dimensional irreducible unitary representation.
So 
there is no plausible copy of adjoint $SU(3)$ with which
to model gluons, although there are representations such as $\rep3b+\rep5$ that might be a suitable substitute.

Therefore the proposed model is inconsistent with quantum chromodynamics (QCD) as a theory of the strong force.
That does not mean it is inconsistent with experiment. Nor does it mean that it is necessarily inconsistent with lattice QCD.
By working on a lattice, one reduces the symmetry to a finite group. If the lattice that is chosen is an ordinary cubic lattice
then the symmetry is reduced to the subgroup $SL(2,3)$ of $SL(2,5)$. Then the representations restrict as follows:
\begin{align}
\rep3a+\rep3b+\rep4a+\rep5&\rightarrow 3+3+(1+3)+(2+3),\cr
\rep1+\rep4a+\rep5+\rep5&\rightarrow 1+(1+3)+(2+3)+(2+3).
\end{align}
These two are still not the same, so there still appears to be some inconsistency between the proposed model and lattice QCD.
However, the difference may be subtle, and may be difficult to distinguish experimentally. Since this discussion takes us beyond the
standard model, I postpone it to Section~\ref{BSM}.

\subsection{Gell-Mann matrices}
In order to obtain the finite analogue of the Gell-Mann matrices we have to work with the representation
$\rep3a\otimes\rep3b=\rep4a+\rep5$. This representation can be written in terms of $3\times3$ matrices, with $\rep3a$ acting by
right-multiplication by the matrices already given, and $\rep3b$ acting by left-mutiplication by the transposed matrices. 
Calculation then shows that $\rep4a$ is spanned by the matrices
\begin{align}
\label{4abas}
\begin{pmatrix}1&0&0\cr 0&1&0\cr 0&0&1\end{pmatrix},
\begin{pmatrix}0&-\sigma&0\cr \tau&0&0\cr0&0&0\end{pmatrix},
\begin{pmatrix}0&0&\tau\cr0&0&0\cr-\sigma&0&0\end{pmatrix},
\begin{pmatrix}0&0&0\cr0&0&-\sigma\cr0&\tau&0\end{pmatrix},
\end{align}
while $\rep5$ is spanned by the matrices
\begin{align}
\begin{pmatrix}1&0&0\cr 0&1&0\cr 0&0&-2\end{pmatrix},
\begin{pmatrix}1&0&0\cr 0&-1&0\cr 0&0&0\end{pmatrix},
\begin{pmatrix}0&\sigma&0\cr \tau&0&0\cr0&0&0\end{pmatrix},
\begin{pmatrix}0&0&\tau\cr0&0&0\cr\sigma&0&0\end{pmatrix},
\begin{pmatrix}0&0&0\cr0&0&\sigma\cr0&\tau&0\end{pmatrix}.
\end{align}
Then the Gell-Mann matrices can be obtained as suitable complex linear combinations of these, using the relations
\begin{align}
\begin{pmatrix}0&1\cr 1&0\end{pmatrix}&=\frac{\sqrt{5}}{2}\begin{pmatrix}0&-\sigma\cr\tau&0\end{pmatrix}
-\frac12\begin{pmatrix}0&\sigma\cr\tau&0\end{pmatrix},\cr
\begin{pmatrix}0&-1\cr 1&0\end{pmatrix}&=\frac{\sqrt{5}}{2}\begin{pmatrix}0&\sigma\cr\tau&0\end{pmatrix}
-\frac12\begin{pmatrix}0&-\sigma\cr\tau&0\end{pmatrix}.
\end{align}

The basis chosen here clearly breaks the symmetry of $\rep4a$ to $1+3$, and of $\rep5a$ to $1+1+3$, but there are more
symmetrical spanning sets that exhibit the structures as deleted permutation representations on $5$ and $6$ points respectively.
Thus in $\rep4a$ we can choose five matrices adding to zero, and permuted by the group:
\begin{align}
\begin{pmatrix}1&0&0\cr 0&1&0\cr 0&0&1\end{pmatrix},
\frac14\begin{pmatrix}-1 & 2+\sigma & 2+\tau\cr 2+\tau & -1 & 2+\sigma\cr 2+\sigma&2+\tau & -1\end{pmatrix},
\frac14\begin{pmatrix}-1 & -2-\sigma & -2-\tau\cr -2-\tau & -1 & 2+\sigma\cr-2-\sigma&2+\tau & -1\end{pmatrix},\cr
\frac14\begin{pmatrix}-1 & -2-\sigma & 2+\tau\cr -2-\tau & -1 & -2-\sigma\cr 2+\sigma&-2-\tau & -1\end{pmatrix},
\frac14\begin{pmatrix}-1 & 2+\sigma & -2-\tau\cr 2+\tau & -1 & -2-\sigma\cr -2-\sigma&-2-\tau & -1\end{pmatrix}.
\end{align}
Here the relations $2+\tau=\sqrt5\tau$ and $2+\sigma=-\sqrt5\sigma$ may be useful for understanding the structure.
The five matrices displayed here correspond to the five quaternions
\begin{align}
1, (-1+\sqrt5(\ii+\jj+\kk))/4,(-1+\sqrt5(\ii-\jj-\kk))/4,\cr
(-1+\sqrt5(-\ii+\jj-\kk))/4,(-1+\sqrt5(-\ii-\jj+\kk))/4.
\end{align}
It follows from this that the basis elements for $\rep4a$ given in (\ref{4abas}) correspond to the quaternions
$1,\ii,\jj,\kk$ in some order, which we can choose arbitrarily. % respectively.

Similarly, in $\rep5$ the following six matrices are permuted by the group:
\begin{align}
&\begin{pmatrix}0&0&0\cr 0&1&\tau\cr 0&\sigma & -1\end{pmatrix}, 
\begin{pmatrix}-1&0&\sigma\cr 0&0&0\cr \tau&0&1\end{pmatrix},
\begin{pmatrix}1&\tau&0\cr \sigma & -1 & 0\cr 0&0&0\end{pmatrix},\cr
&\begin{pmatrix}0&0&0\cr 0&1&-\tau\cr 0&-\sigma & -1\end{pmatrix}, 
\begin{pmatrix}-1&0&-\sigma\cr 0&0&0\cr -\tau&0&1\end{pmatrix},
\begin{pmatrix}1&-\tau&0\cr -\sigma & -1 & 0\cr 0&0&0\end{pmatrix}.
\end{align}
Note, incidentally, that these matrices have rank $1$ rather than $2$. 
Hence they can be factorised as the product of a row vector (in $\rep3a$) and a column vector (in $\rep3b$),
but not uniquely, since scalars can be transferred from one vector to the other.

Now it is easy to imagine these six matrices coming in three pairs, obtained by changing the sign on the off-diagonal
entries, and it is easy to imagine this pairing to be obtained by complex conjugation. One can then convert the
representation $\rep5$ into a $3$-dimensional unitary representation of $SU(3)$. But this imposes a structure that
is incompatible with the finite symmetry group, and creates a structure of three colours and three anticolours
that span a $5$-space, not a $6$-space. In other words, the finite model has colour confinement built into
it naturally. At the same time, we are able to match the permutation representation $\rep1+\rep5$ to the monomial representations
$\rep3a+\rep3b$ and $\rep6$, in order to investigate how this apparent breaking of symmetry arises from
the underlying algebraic structure.

There is a third permutation representation
worth considering here: $\rep4a+\rep5$ is the deleted permutation representation on $10$ points,
which can be taken to be the following matrices:
\begin{align}
\begin{pmatrix}1&1&1\cr1&1&1\cr1&1&1\end{pmatrix},
\begin{pmatrix}1&-1&-1\cr-1&1&1\cr-1&1&1\end{pmatrix},
\begin{pmatrix}1&-1&1\cr-1&1&-1\cr1&-1&1\end{pmatrix},
\begin{pmatrix}1&1&-1\cr1&1&-1\cr-1&-1&1\end{pmatrix},\cr
\begin{pmatrix}0&0&0\cr0&-1&\tau^2\cr 0&\sigma^2&-1\end{pmatrix},
\begin{pmatrix}-1&0&\sigma^2\cr0&0&0\cr \tau^2&0&-1\end{pmatrix},
\begin{pmatrix}-1&\tau^2&0\cr\sigma^2&-1&0\cr 0&0&0\end{pmatrix},\cr
\begin{pmatrix}0&0&0\cr0&-1&-\tau^2\cr 0&-\sigma^2&-1\end{pmatrix},
\begin{pmatrix}-1&0&-\sigma^2\cr0&0&0\cr -\tau^2&0&-1\end{pmatrix},
\begin{pmatrix}-1&-\tau^2&0\cr-\sigma^2&-1&0\cr 0&0&0\end{pmatrix}.
\end{align}
These matrices also have rank $1$, so can be factorised as a product of a row vector and a column vector.
In all of these representations, there is a $\sqrt{5}$ that is capable of holding the same discrete quantum numbers
that are assigned to $\sqrt{-1}$ in the standard model. Of course the algebra is quite different, so that translating from
one to other cannot be done without a great deal of `mixing' in order to transfer the quantum numbers, or other variables, 
from one model to
the other. 

\subsection{Mixing}
In the finite model these $3\times 3$ matrices do not generate a Lie algebra or a Lie group, as they do in the
standard model, but simply form a
representation of the finite group. The $\rep4a$ component then appears not only in the strong force, but also in the
electro-weak force, in such a way that the mixing defines four mixing angles, associated to a suitable basis of the representation.
These are presumably the mixing angles in the  Cabibbo--Kobayashi--Maskawa (CKM) matrix 
\cite{Cabibbo,KM}, consisting of an overall `CP-violating phase' associated to
the identity matrix, together with the Cabibbo angle and two other generation-mixing angles. 

There are, of course, five other parameters here that correspond to the representation $\rep5$. Four of these presumably 
appear in  the Pontecorvo--Maki--Nakagawa--Sakato (PMNS) matrix
\cite{Pontecorvo,MNS} in an analogous way, while the fifth must be the electro-weak mixing angle. A precise allocation of
these mixing angles to the matrices in the finite group algebra must await further development of the model. It may be
that the CKM and PMNS matrices represent maps between $\rep4a$ and $\rep5$, rather than vectors within them, or perhaps
I have allocated them the wrong way round.

\section{Beyond the standard model}
\label{BSM}
\subsection{A proposal}
It has become clear from the preceding discussion that a clear distinction between $\rep2a+\rep2b$ and $\rep4b$ needs to be made.
The former is acted on by $Spin(4)$, the latter by $Spin(5,1)$. The Lorentz group can be obtained as a subgroup $Spin(3,1)$
of $Spin(5,1)$. Therefore $\rep2a+\rep2b$ is a Lorentz-invariant spinor, while $\rep4b$ is a Lorentz-covariant spinor. 
In theory, the standard model describes Lorentz-invariant spinors as `isospinors', which are discrete internal variables.
But in practice, 
the experimentally discrete electron generation symmetry is conflated with the experimentally
continuous neutrino generation symmetry.
This is a clear sign that that Lorentz-covariant symmetries 
have been conflated with
Lorentz-invariant symmetries.

I therefore suggest that we take $\rep4b$ as the Lorentz-covariant spinor, so that we can extend from $Spin(3,1)$ to
$Spin(5,1)$ to take account of the different Lorentz groups chosen by different observers, who each have their own
preferred definition of inertial frame. 
This will allow us to distinguish between a particle physics inertial frame
attached to the Earth, and a gravitational inertial frame which is not. It will also allow us to distinguish between a
Solar System inertial frame, appropriate for  Newtonian gravity and general relativity, and a galactic inertial frame,
appropriate for a more general theory of gravity. 

On this basis we can treat the gauge groups of the theory as consisting of
\begin{itemize}
\item $Spin(4)=SU(2)_L\times SU(2)_R$ acting on the invariant spinor;
\item $Spin(5,1)=SL(2,\HH)$ acting on the covariant spinor; and
\item $SL(3,\HH)$ acting on the rest of the model.
\end{itemize}
From here we can restrict to
\begin{itemize}
\item $SU(2)_L$ as the gauge group of the weak interaction;
\item $Spin(3,1)=SL(2,\CC)$ as the Lorentz group; and
\item $SU(3)$ inside $SL(3,\CC)$ as the gauge group of the strong force, and a scalar $U(1)$
as the gauge group of electrodynamics.
\end{itemize}
Hence we should be able to recover the standard model readily enough as a sub-model of the proposed new model,
although the discussion in the previous section suggests there may be a difficulty with implementing QCD. 

Moreover, we have a group $SU(2)_R$ which can act as a generation symmetry group, in order to extend the
standard model to three generations of fermions.
In addition, we obtain a gauge group which might possibly be useful 
for a quantum theory of gravity, by restricting to the compact subgroup
$Spin(5)$ of $Spin(5,1)$. Finally, 
the compact subgroup $Sp(3)$ of $SL(3,\HH)$ is $21$-dimensional, and contains some version
of the gauge group $SU(3)$ of the strong force.
I propose that it might generalise the $20$-dimensional Riemann Curvature Tensor, by allowing the invariant
scalar (mass) to become covariant instead.

Now to model the forces, we need to map from the gauge groups acting on the spinors, to corresponding
groups acting on the faithless part of the algebra. This puts the weak force in $\rep3a$ and the generation symmetry in $\rep3b$,
as well as putting gravity in $\rep5$. But there is no similar invariant action of $SU(3)$ or $U(1)$, so that in these cases
the translation from gauge group to bosons cannot proceed in the standard Yang--Mills fashion. Nevertheless, I propose
that all the four fundamental forces can be modelled in some way inside $\rep3a+\rep3b+\rep4a+\rep5$. In the standard model they
mix together in various complicated ways, that are covariant rather than invariant, since they are experimentally
known to vary
with the energy. In the group algebra model, however, it is possible to split into four forces in an \emph{invariant} way.
Clearly, therefore, this is not exactly the same as the standard model splitting, but, if it works, it should produce a much simpler model
in the end.

Mediators for invariant forces must presumably be massless, and they must be bosonic as far as the Lorentz group is concerned. 
Of course, a Lorentz-invariant spinor is,
counter-intuitively, bosonic in this sense. 
In particular, elements of $\rep2a+\rep2b$ are physically bosons, although mathematically 
fermionic in this model.
In other words, neutrinos are technically bosons. This is a consequence of
our separation of the two concepts of Lorentz-invariant and Lorentz-covariant spinors.  
Therefore, neutrinos are
available as mediators for both quantum gravity and the invariant version of the weak force.
Of course, this means that the Yang--Mills paradigm of quantum field theory does not apply, 
but that was always going to be the case for a discrete model, so cannot be taken as a criticism.
We can now interpret $\rep3a+\rep3b$ as photons and $\rep3a\otimes \rep3b=\rep4a+\rep5$ as neutrinos, for a complete set
of mediators. 

This suggestion may change our perspective on whether to regard neutrinos as Lorentz-covariant
or Lorentz-invariant entities.
It is not quite as obvious as it might seem that the covariant symmetries belong to the neutrino,
and the invariant symmetries to the electron, since the experimental evidence relates mainly to a mixing of the two.
The mathematics here suggests that it may be useful to
convert from the standard model Lorentz-covariant
(and therefore massive) neutrinos to Lorentz-invariant (and therefore massless) neutrinos.

\subsection{Hamiltonian formulation}
\label{Lagrange}
In order to formulate a Hamiltonian, we need to identify a particular representation to hold potential and kinetic energy.
The obvious choice is $\rep4a$, of which there are four copies in the group algebra. Now we can 
separate 
the four copies 
by  tensoring $\rep4a$ with each of $\rep3a$, $\rep3b$, $\rep4a$ and $\rep5$ in turn, and projecting onto the unique copy of $\rep4a$ in each case.
The tensor product here corresponds to multiplication or division by an infinitesimal, so that the whole operation can be
converted into integration or differentiation if desired. 

At the macroscopic level, we have four copies of $\rep4a$ in the Hamiltonian, and they can be mixed together in arbitrary ways, with a 
$GL(4,\mathbb R)$ gauge group. 
That is the group of general covariance, that says we can split the Hamiltonian any way we like, 
and we will always get the same answer. The splitting of the Hamiltonian that is adopted in the standard model is 
therefore completely arbitrary, and the so-called constants are 
really artefacts of our decision to separate gravity from the other forces, 
and therefore they are artefacts of the particular gravitational environment we live in. 
The finite group suggests a completely different way to split the Hamiltonian, in such a way that
we separate four quantised forces rather than three, and therefore have a quantum gravity that 
consists of half of the strong force, together with one more dimension that is new.

We then obtain the Hamiltonians from consideration of the tensor products
\begin{align}
\rep3a\otimes \rep4a &= \rep3b+\rep4a+\rep5\cr
\rep3b\otimes \rep4a &= \rep3a+\rep4a+\rep5\cr
\rep4a\otimes \rep4a &= \rep1+\rep3ab+\rep4a+\rep5\cr
\rep5\otimes \rep4a &= \rep3a^2b^2+\rep4a+\rep5.
\end{align}
The first two lines contain the Hamiltonian for quantum electrodynamics, with the clear action of the Dirac matrices in $\rep4a$
swapping the left-handed $\rep3a$ with the right-handed $\rep3b$. The third line contains the modified Dirac algebra, with a clear
distinction between $\gamma_5$ in the $\rep5$ representation, and the scalars in $\rep1$. While these three lines in the standard model
are used only for electro-weak unification, it is clear at this point that they contain half of the strong force as well.
The other half of the strong force is in the last line, and differs from the third line in having $\rep3a+\rep3b$ (three colours and three anti-colours)
in place of $\rep1$ (no colour), thereby enforcing colour confinement.

The usual interpretations of $\rep3a+\rep3b$ as the electromagnetic field, and as photons, regarded as spin $1$ particles, are still viable in this
scenario, in which $\rep3a$ and $\rep3b$ are the spin $1$ (adjoint) representations of $SU(2)_L$ and $SU(2)_R$ respectively. But $\rep4a$ is actually a
spin $1/2$ representation of both $SU(2)_L$ and $SU(2)_R$. Hence an interpretation as spin $1$ mediators is not viable in this case.
The only reasonable interpretation of the group algebra model is then that the representation $\rep4a$ consists of neutrinos and/or antineutrinos.
The representation $\rep5$, on the other hand, is acted on by $Spin(5)$ rather than $Spin(4)$. If we mix the two together by embedding $Spin(4)$
in $Spin(5)$, we split $\rep5$ into $1+4$, and must interpret these pieces as spin $0$ and spin $1/2$, rather than spin $2$.
If, however, we look at the action of $Spin(3,1)$, we see a splitting $1+1+3$, interpreted as two spin $0$ and three spin $1$ particles,
but then presumably mixed together into a set of five spin $1$ gluons.

\subsection{Some implications}
Another way of looking at this proposal is to think of the representation $\rep4a+\rep5$ as consisting of gluons. This interpretation
is certainly more conventional, but it requires there to be $9$ gluons rather than $8$. The ninth gluon could then be interpreted
as a graviton. But it is not well-defined. A better interpretation might be to consider $\rep3a$ and $\rep3b$ to be 
the momenta of neutrinos and antineutrinos respectively, so that $\rep3a+\rep3b$ describes a photon obtained from the
annihilation of a neutrino and an antineutrino. Then $\rep3a\otimes \rep3b$ must describe an interaction in which a nuclear process
emits a neutrino and/or an antineutrino.

 In the standard model, the weak force accounts for $4$ such processes, one of which
(beta decay)
emits an antineutrino, while the other $3$ (change of electron generation) emit both. The other $5$ are then strong force processes,
in which the combination of neutrino and antineutrino is interpreted as a gluon. 
This creates an overlap between the $3$ generation-changing processes of the weak force, and
the other three gluons, and is responsible for the weak-strong mixing within the standard model.
It also creates an overlap between beta decay and the proposed ninth gluon, or graviton, and hence creates a
mixing between the weak force and gravity, that gives mass to the intermediate vector bosons.
The corresponding part of the standard model is a set of four massive bosons, consisting of the three $Z$ and $W$ bosons
plus the Higgs boson, forming a basis for another copy of $\rep4a$.

A further consequence of the separation of invariant from covariant spinors is that the particle-antiparticle distinction becomes a 
distinction between the two halves of the invariant spinor, rather than the covariant spinor. Hence the proposed model
does not permit an interpretation of an antiparticle as being a particle travelling backwards in time.
In this model, there is no CPT-symmetry, and 
a particle-antiparticle pair is better interpreted as a C-doublet rather than a PT-doublet. 
It follows that this model contains an intrinsic (invariant) distinction between particles and antiparticles, such that there is
no particle-antiparticle symmetry. In other words it explains the particle-antiparticle asymmetry in the universe.

Yet another consequence is the discreteness of photon polarisation, since the photon lies in the representation $\rep3a+\rep3b$,
and the two components represent the two polarisations, that is, the two helicities in the standard terminology.
But by using $SO(4)$ rather than $SO(3,1)$, we convert the helicity into a chirality, and a Lorentz-covariant angular momentum
into a Lorentz-invariant momentum with an additional sign for the chirality.
Linear polarisation arises then from interactions with matter, which sets up a correlation between the internal symmetries in $\rep3a+\rep3b$
and the environmental symmetries in $\rep4b$. The measurement problem arises from the identification of $\rep2a+\rep2b$ with $\rep4b$,
which causes an identification of $\rep3a+\rep3b$ with the adjoint representation of the Lorentz group, and therefore an identification
of an observer-dependent property with an observer-independent property. 
In other words, the measurement problem would seem to be caused, in effect, by the confusion between $SO(4)$ and $SO(3,1)$ that 
pervades the whole Dirac algebra, 
and the resulting
confusion between internal properties and environmental properties.

\subsection{Quantum gravity}
\label{mass}
As has already been said,
I take it as fundamental that the symmetry group of Einstein's equation
\begin{align}
m^2c^4+p^2c^2-E^2&=0
\end{align}
is the group $SO(4,1)$. At the quantum level it is necessary to restrict to a compact group, that is $SO(4)$,
which implies that energy is treated as a scalar.
In relativity, on the other hand, it is usual to treat (rest) mass as a scalar, and therefore to work with the group $SO(3,1)$.
The way that these groups are combined in the usual formalism is to complexify the Clifford algebra, so that both signatures can
be treated simultaneously. However, this blurs the distinction between mass and energy in a rather unfortunate way.

It seems to me, therefore, that it may be better to maintain a clear distinction between mass and energy, and work with $5$-momentum
rather than $4$-momentum. Since $5$-momentum is a generalisation of both energy and mass, it would seem to be a natural
choice for the fundamental concept
on which any realistic quantum theory of gravity must ultimately be based. The gravitational force between two bodies defined by their mutual
$5$-momenta must, by Newton's third law, be described by an anti-symmetric tensor. 
Since Newton's law of universal gravitation
depends on the product of the masses, this tensor has rank $2$. It is therefore a tensor with $10$ degrees of freedom,
the same number as there are in both the stress-energy tensor and the Ricci tensor in general relativity.

The difference, however, is that because general relativity only works with $4$-momentum, not $5$-momentum, the 
two tensors are 
defined instead as symmetric rank $2$ tensors on a Minkowski $3+1$ spacetime, or its dual. While such a tensor can fulfil many of the
same functions as the anti-symmetric tensor on $5$-momentum, it is not the same tensor. 
Moreover, since a symmetric tensor does not comply with Newton's third law,
the basic premise of general relativity implies that gravity is not a force in the Newtonian sense. That is, of course, the usual modern
interpretation. But if it isn't regarded as a force, then it is hard to see how it can 
be quantised in any reasonable way.

In particular, the spin $2$ representations that appear in the Ricci tensor restricted to $SO(4)$, and in the Riemann curvature tensor
restricted to $SO(3,1)$, cannot in this scheme represent quantised gravitons. The gravitons must live in the anti-symmetric tensor, which breaks up as
$3+3+4$ for $SO(4)$. Here, $3+3$ represents spin $1$ particles, namely photons, and $4$ represents spin $1/2$ particles, namely
neutrinos. This contrasts with the Ricci tensor, which breaks up as $1+4+5$ for $SO(4)$, and which may well represent 
physical particles, although it is hard to see how these could be interpreted as 
gravitons.

Now we are ready to consider how to do the quantisation. For this purpose, we need to use a finite group, and I propose that $Alt(5)$
is the most suitable group, for all the many reasons I have already given. The permutation representation on $5$ letters breaks up
as $\rep1+\rep4a$, on which there is an invariant Lorentzian metric. However, this does not permit any mixing of the scalar, whether that be
interpreted as energy or mass, with the rest of the representation. A better option, therefore, may be to consider
using  the irreducible
representation $\rep5$, that is the monomial representation, that puts complex scalars of order $3$ on each of the five letters.
This has the effect that the mass term comes as a triplet, for three generations, and so do the momentum 
(or massless neutrino) terms.

On the basis of these hypotheses, we conclude that quantum gravity must be described by the tensor
\begin{align}
\Lambda^2(\rep5)&=\rep3a+\rep3b+\rep4a\cr
&=\Lambda^2(\rep1+\rep4a).
\end{align}
In particular, the change from the permutation representation to the monomial representation does not change the
analysis of the gravitational force. But it does change the analysis of mass and energy, and incorporates the $3$-generation
structure of matter directly into the definition of (gravitational) mass. Hence it provides a modification of Newtonian gravity that is a
consequence of the existence of three generations. This 
seems to imply a failure of the equivalence principle, which will be discussed further in Section~\ref{EP}.

\section{Predictions and postdictions} 
\subsection{A new quantum process?}
\label{newprocess}
The standard model contains $6$ degrees of freedom for the photon, plus $8$ degrees of freedom for gluons, 
making a total of $14$ for the massless force mediators. By allowing the neutrinos into the picture, I extend this to $15$.
In other words, the model predicts an extra quantum process that is mass-neutral, and that does not exist in the standard model.
I have already made such a prediction elsewhere, namely the (hypothetical) process
\begin{align}
e+\mu+\tau +3p &\leftrightarrow \nu + 5n.
\end{align}
Certainly the total masses on the two sides are equal to within experimental uncertainty, as is the total charge. 

It is not clear whether a single neutrino is sufficient or whether perhaps three are required,
but this makes no material difference to the argument, and does not change any of the conclusions. 
If this process takes place right-to-left, then the $\mu$ and $\tau$
will then decay to electrons with the release of a vast amount of energy in the form of heat and neutrinos and antineutrinos (but no
photons, so the process is invisible). In order to get
the neutrons close enough together for this reaction to take place, they must be inside the nucleus of an atom. And in order for the $\mu$ and $\tau$
to get far enough away to avoid recombination, there must either be very large tidal forces on the atomic nucleus, or the atom must be highly ionised,
or both. 

If this process is combined with an inverse $\beta$ process, then we obtain 
\begin{align}
4n&\rightarrow \mu+\tau+2p.
\end{align} 
This removes the 
need for the neutrino, but now the mass on the right hand side is slightly greater than that on the left hand side,
so the process needs an energy input to get going. But then it can take place spontaneously in a very hot, completely ionised beryllium
atom, rotating very fast in a strong gravitational field, due to the presence of a strong magnetic field. Such conditions exist in the
solar corona, so that this process provides a plausible mechanism for solar coronal heating. At the same time, it burns up two whole nucleons,
and tries to create a nucleus of nothing but protons, which of course flies apart into hydrogen nuclei, or perhaps it can be combined with 
an inverse $\beta$ process to leave some helium behind.

In the opposite direction, the process creates nucleons out of leptons, but needs a far greater energy input, and therefore much more extreme
conditions, such as centres of stars, or the Big Bang. But it is a process of nucleosynthesis that is unknown to the standard model,
and if it takes place in such extreme conditions then it might explain why the standard model predictions of metal abundances in the
universe do not appear to be consistent with observations. In particular, the process could take place with the three electrons and three protons in
a lithium atom, under conditions of extreme pressure and extreme heat, by first converting the electrons to the higher generations,
and then converting the whole lot into neutrons. Thus the original $7$ nucleons in lithium-7 have been converted into $9$,
and after some beta decay to create electrons and protons, one ends up with beryllium-9.
In particular, this process might be involved in the formation of neutron stars. Moreover, since it takes out such a huge amount of energy,
it results in rapid cooling.

\subsection{
Reduction to general relativity}
\label{GR}
The proposal here for quantum gravity is to base it on the representation $\rep5$, rather than $\rep4a$
as in general relativity.
Effectively, then, we extend the field-strength tensor 
\begin{align}
\Lambda^2(\rep4a)&=\rep3a+\rep3b
\end{align}
of general relativity to the larger tensor
\begin{align}
\Lambda^2(\rep5)&=\rep3a+\rep3b+\rep4a. 
\end{align}
This suggests that the stress-energy tensor should similarly be extended from 
\begin{align}
S^2(\rep4a)&=\rep1+\rep4a+\rep5
\end{align} 
to 
\begin{align}
S^2(\rep5)&=\rep1+\rep4a+\rep5+\rep5,
\end{align}
that is by adding an extra copy of $\rep5$. 

Therefore we extend the $10$ Einstein equations to $15$. These equations come from
the equivalence of representations
\begin{align}
S^2(\rep5)&\cong\Lambda^2(\rep6).
\end{align}
The extra $5$ equations appear to involve $5$ fundamental gravitational masses, so that mass itself is a $5$-vector in this model,
rather than a scalar. Rather than taking only the neutron, proton and electron as fundamental, and equating the masses of neutron and
proton more or less to $1$ and the electron to $0$, for a scalar mass, we seem to need to take these three masses independently,
augmented by the other two generations of electron. The extra five equations then relate these five masses to properties of the spin $2$
part of the gravitational action.

Let us now consider the Riemann Curvature Tensor. As it stands, this tensor corresponds to
\begin{align}
S^2(\Lambda^2(\rep2a\otimes \rep2b))&=S^2(\Lambda^2(\rep4a))\cr
&=S^2(\rep3a+\rep3b)\cr
&=\rep1^2+\rep4a+\rep5^3.
\end{align}
But this tensor does not take account of the fermionic nature of the electron, proton and neutron. 
It may be worthwhile
to replace the `bosonic' monomial representation $\rep3a+\rep3b$ by the corresponding `fermionic' representation $\rep6$, thus:
\begin{align}
S^2(\rep2a\otimes S^2(\rep2b))&=S^2(\rep2a\otimes \rep3b)\cr
&=S^2(\rep6)\cr
&=\rep3a^2b^2+\rep4a+\rep5.
\end{align}

The effect of this proposed change is to replace two copies of $\rep1+\rep5$, which might represent the particles $\nu+5n$, by $\rep3a+\rep3b$,
which might represent $e+\mu+\tau+3p$. Since this does not change the overall mass, it has no effect on Newtonian gravity,
but it does have an effect on general relativity. Most notably, it is no longer possible to take out a scalar from the $21$-dimensional
tensor in order to leave a $20$-dimensional curvature tensor. This scalar now plays the role of breaking the $3$-generation symmetry,
and picking out the first generation electron as special. But the group algebra requires this symmetry to be unbroken.
In other words, this modification of the Riemann Curvature Tensor is \emph{required} in order to take account of the fact that there are
three generations of electron.

This 
proposed modification 
of the Riemann Curvature Tensor
corresponds to the adjoint representation of the gauge group $Sp(3)$, that has not been used at all so far.
It therefore corresponds to a group of coordinate changes, that 
appears to describe changes between different observers' experiences of \emph{matter}.
Moreover, it contains a copy of $SU(3)$, suitable for describing our experiences of the internal structure of the
proton and the neutron.

We have now split up the entire faithless part of the group algebra, into a field strength tensor $\Lambda^2(\rep5)\cong S^2(\rep4b)$,
two equivalent
representations $S^2(\rep5)$ and $\Lambda^2(\rep6)$ sharing a scalar, whose equivalence contains and extends the Einstein equations,
and $S^2(\rep6)$, which 
enhances the Riemann Curvature Tensor as a description of the `shape of spacetime' as
defined by the \emph{observer's assumptions about} the matter within it, plus one extra dimension to define the \emph{local} mass scale. 

\subsection{In search of the equations}
\label{equations}
If the analysis in Section~\ref{GR}
 is correct, then it follows that the fundamental mass ratios must be related to properties of the rotation of the Earth.
 Clearly this cannot imply that there is any 
\emph{physical} change in mass if the rotational and/or gravitational parameters change. What it must mean is that our historical
\emph{choice} of gauges for the theory has been determined by our particular place in the universe.

The required properties of the Earth's motion 
can be
roughly separated into components defined by its rotation on its own axis, its revolution around the Sun, and the revolution of the Moon
around the Earth. Of course, these components are not completely independent of each other, so it will be impossible to write down any
exact equations at this stage. But we should be able to find some equations that express the five fundamental masses in terms of
rotational and gravitational parameters, to an accuracy of better than $1\%$. 

If we neglect the gravitational effects of the Moon for the moment, there are just two obvious dimensionless parameters available, which we might as well
take to be the number of days in a year, say $365.24$, and the tilt of the Earth's axis, say $23.44^\circ$. With these parameters, we want to calculate
the mass ratios of three particles, presumably $e$, $p$ and $n$. Experimentally, we have
\begin{align}
n/p&\approx 1.001378,\cr
e/p&\approx .000544617,
\end{align}
for which the only plausible formulae are
\begin{align}
1+1/(2\times 365.24)&\approx 1.001369,\cr
\sin(23.44^\circ)/(2\times 365.24) &\approx .000544558.
\end{align}
Hence both formulae are accurate to within about $.01\%$, which is much more accurate than we could reasonably have hoped.
Even the \emph{difference} between the neutron and proton masses is postdicted to an accuracy of better than $1\%$.

Of course, this does \emph{not} mean that these mass ratios change if the tilt of Earth's axis changes, or the number of days in a year changes.
That would be absurd, and is in any case contradicted by experiment. What it means is that the way that the theories have developed, and the way
that various choices of coordinates have been made, reveal a dependence of sorts on the environment we live in. 
What it means is that the structure of the model that we have developed depends on what we feel is important to measure.
What it means is that
certain parameters that can be treated as constants in the standard model, do not necessarily have to be treated as constants.
A more general model in which they are treated as variables may then reveal things about the universe that are not revealed by 
the standard model. 

\subsection{Pions and kaons}
\label{pionkaon}
At this stage we have two more gravitational parameters, but only one more mass ratio, so perhaps I have not made the right choice of
fundamental mass ratios. 
Here I  
consider an alternative. 
The change of signature from $SU(3)$ to $SU(2,1)$ imposed by the structure of the group algebra has profound consequences for the
interpretation of the strong force. Instead of the $8$ massless gluons that are assumed to lie in the adjoint representation of $SU(3)$, we
now appear to have $8$ gauge bosons which between them have four distinct masses. 
These must include the three pions, that mediate the strong force
between 
nucleons, 
and presumably also the kaons. This gives us 
$4$ masses, but 
seemingly 
only $7$ particles.

However, it is clear that the four boosts must represent the four charged particles, so that the rotations represent the neutral particles,
forming a group $U(2)$, and therefore splitting $1+3$ into mass eigenstates. In other words, the model counts three neutral kaons,
in the adjoint representation of $SU(2)_R$, rather than two, in two separate $2$-dimensional representations of $SU(2)_L$, as
in the standard model. Hence the group algebra model is incompatible with the standard model at this point. However, it is 
not incompatible with experiment, which very clearly detects three distinct neutral kaons with the same mass.
Moreover, the model provides a physical explanation for kaon oscillations between the three eigenstates, in terms of the action of
$SO(3)_R$, which represents the magnetic and/or gravitational field.

That is, a kaon changes state as a result of an interaction with a magnetic field associated with quantum gravity.
In practical terms, therefore, it is an interaction with
an antineutrino. These antineutrinos combine in pairs (one in and one out) to create the gravitational field in $\rep3b$. In other words,
the source of the antineutrinos that cause kaon oscillations is the gravitational field itself. Hence the `mixing angle' between
kaon states is simply the angle between the directions of the gravitational field at the locations of the two measurements.

The original experiment \cite{CPexp} that detected these oscillations was carried out over a horizontal distance
of $57$ feet, that corresponds to an angle of approximately $2.73\times 10^{-6}$ radians, assuming a mean radius of the Earth
of around $6367$ km. The lifetimes of the relevant two kaon eigenstates differ by a factor of approximately 570, which means that
the model postdicts that approximately $.16\%$ of detected decays will be two pion decays.
The experiment found $45\pm9$ in a sample of $22700$, that is $.20\%\pm.04\%$, so that my postdiction is consistent with the experimental
results, to an accuracy of $1\sigma$.

Admittedly, it is only a postdiction at this stage, but it can be converted into a genuine prediction that the effect
depends on the geometry of the experiment relative to the gravitational field. Hence I predict that different results will be obtained
for different experimental geometries. This can be tested. Given that the standard model makes predictions about kaon decays
that are not consistent with experiment \cite{kaonanomaly,Kaon2}, I suggest that my alternative proposals should be tested.

Now it must also be taken into account that in order to obtain the subgroup $SU(2,1)$ of $Spin(5,1)$ we have to break the symmetry of the finite
group, from the binary icosahedral group $SL(2,5)$ to the binary tetrahedral group $SL(2,3)$. This breaks the symmetry of the
representation $\rep5$ into $2+3$, in such a way that the $3$ is equivalent to the restrictions of both $\rep3a$ and $\rep3b$. Hence the
$3$-dimensional representation is now playing three roles simultaneously, in the weak force, the strong force and gravity.

Going back to the unbroken symmetry, we have both pion and kaon masses represented in the representation $\rep5$, and hence we can
hope to find mass ratios corresponding to the number of days in a month, and the inclination of the Moon's orbit. For this purpose,
a lunar month seems most appropriate, at $29.53$ solar days, but we should not expect very great accuracy from any particular
way of counting the number of days in a month. The inclination of the Moon's orbit is around $5.14^\circ$ on average, but varies between
$4.99^\circ$ and $5.30^\circ$ on a cycle of length approximately $343$ days, so again we should not expect great accuracy.

Experimental values of the mass ratios are
\begin{align}
\pi^\pm/\pi^0&\approx 139.570/134.977\cr
&\approx 1.03403\cr
K^\pm/K^0&\approx 493.68/497.65\cr
&\approx .99202
\end{align}
compared to possible theoretical estimates
\begin{align}
1+1/29.53&\approx 1.03386\cr
\cos^2(5.14^\circ)&\approx .99197,
\end{align}
where the appearance of $\cos^2$ arises from the fact that $\rep5$ is a spin $2$ representation.
Of course, these conjectures are speculative, and may not be very convincing until a more complete model is built.
But they are accurate to $.05\%$, which would be quite remarkable if these were pure coincidences.

\section{The equivalence principle}
\label{EP}
\subsection{Gravitational and inertial mass}
The strange formulae for certain mass ratios in the previous section can only hold if the strong equivalence principle fails.
Otherwise, precise measurements of inertial masses over the past 50 years directly contradict the implied
variability of gravitational mass ratios. Experimental evidence therefore requires inertial mass of the particles
under consideration to be fixed, and requires gravitational mass of bulk matter, and therefore of atoms, to be fixed,
within the limits of experimental uncertainty. The standard model permits the calculation of inertial masses of atoms,
at least in some cases, 
in terms of inertial mass of electrons, protons and neutrons, together with
other parameters of the theory. These inertial masses can then be calibrated against gravitational masses.

Direct measurement of gravitational mass is difficult to perform accurately, and depends on an accurate value for
Newton's gravitational constant $G$. This constant is notorious for being difficult to measure, and 
the CODATA 2018 value has a relative standard uncertainty of 22ppm, which may be over-optimistic in view of the
inconsistencies between different experiments. It is certainly not possible to measure the gravitational masses
of individual electrons and protons to anything like this accuracy, so that a direct test of the proposed formulae for gravitational
mass ratios is impossible.
But indirect tests may be possible, particularly if historical measurements or historical theories have in practice
dealt with a unified practical `mass' which is a mixture of a theoretical `inertial mass' and `gravitational mass'.

In order to investigate this possibility, let us define inertial, gravitational and experimental mass ratios for the
proton and electron as follows:
\begin{align}
R_I&:= 1836.15267343\cr
R_G&:= 2\times 365.256363/\sin\theta\cr
R_E&:= R_I\cos^2\lambda + R_G\sin^2\lambda,
\end{align}
where $365.256363$ is the number of solar days in a sidereal year, $\theta$ is the Earth's axial tilt or obliquity,
and $\lambda$ is a mixing angle that may depend on both the type of experiment
and the model that is used to interpret it. The historical record makes it clear that since the consolidation of the
standard model in the 1970s, $\lambda=0$. The question then is whether a different value of $\lambda$
was used before the standard model.

Accurate values of the true obliquity $\theta$ can be
obtained from the Astronomical almanac \cite{almanac}, but approximate values suffice for our
purposes. I tabulate values of 
$\theta$ and $R_G$ for the relevant period here:
\begin{align}
\begin{array}{llllll}
\mbox{Date} & \mbox{Mean} & \mbox{Nutation} & \theta & \sin\theta & R_G\cr\hline
1949&23^\circ26'45.2''&5.7'' & 23^\circ26'50.9''& .3979082 &1835.883\cr
1951&23^\circ26'44.2''&9.0'' &23^\circ26'53.2'' & .3979183 & 1835.835\cr
1953&23^\circ26'43.3''&8.0'' & 23^\circ26'51.3''& .3979099 & 1835.874\cr
1955&23^\circ26'42.4''&3.1'' & 23^\circ26'45.5''& .3978841 & 1835.994    \cr
1957&23^\circ26'41.5''&-3.1'' & 23^\circ26'38.4''& .3978525 & 1836.139   \cr
1959&23^\circ26'40.6''&-8.0'' & 23^\circ26'32.6''& .3978268 & 1836.258   \cr
1961&23^\circ26'39.6''&-9.0'' & 23^\circ26'30.6''& .3978179 & 1836.299   \cr
1963&23^\circ26'38.7''&-5.7'' & 23^\circ26'33.0''& .3978285 & 1836.250   \cr
1965&23^\circ26'37.8''&0 & 23^\circ26'37.8''& .3978499 & 1836.152\cr
1967&23^\circ26'36.9''&5.7'' & 23^\circ26'42.6''& . 3978712 & 1836.053\cr
1969&23^\circ26'36.0''&9.0''& 23^\circ26'45.0''& .3978819 & 1836.004\cr
1971&23^\circ26'35.0''&8.0''& 23^\circ26'43.0''& .3978730 & 1836.045\cr
1973&23^\circ26'34.1''&3.1''& 23^\circ26'37.2''& .3978472 & 1836.164\cr
\hline
\end{array}
\end{align}
In earlier sections I have shown that the electro-weak mixing angle, or Weinberg angle, $\theta_W$ is likely to be
important in the mixing between inertial and gravitational mass. Since experimental measurements of
$\theta_W$ give $\sin^2\theta_W\approx .22$, I calculate the following values of $R_E$
for various values of the mixing angle $\lambda$:
\begin{align}
\begin{array}{l|lllll}
\sin^2\lambda &0 & .2 & .22 & .25 & 1\cr\hline
1949 & 1836.153 & 1836.099 & 1836.093& 1836.085&1835.883\cr
1951 & 1836.153& 1836.089 & 1836.083& 1836.073&1835.835\cr
1953 & 1836.153& 1836.097 & 1836.091& 1836.083 &1835.874\cr
1955 & 1836.153& 1836.121 & 1836.118 &1836.113  &1835.994\cr
1957 & 1836.153& 1836.150 & 1836.150 & 1836.149&1836.139\cr
1959 & 1836.153& 1836.173 &1836.175 & 1836.178&1836.258\cr
1961 & 1836.153& 1836.181 & 1836.184 & 1836.188& 1836.299\cr
1963 & 1836.153& 1836.172 & 1836.174 & 1836.176& 1836.250\cr
1965 & 1836.153& 1836.153 & 1836.153& 1836.153& 1836.152\cr
1967 & 1836.153& 1836.132 & 1836.130 & 1836.127& 1836.053\cr
1969 & 1836.153 & 1836.123 &1836.120 & 1836.116 & 1836.004 \cr
1971 & 1836.153 & 1836.131 & 1836.129 & 1836.126 & 1836.045 \cr
1973 & 1836.153 & 1836.155 & 1836.155 & 1836.156 & 1836.164 \cr
\hline
\end{array}
\end{align}

\subsection{Experimental evidence}
The relevant experiments are analysed and evaluated in the 1969 CODATA adjustment \cite{1969}
and in the 1973 adjustment \cite{1973}. In both cases, the critical experiments were those that
measured the proton magnetic moment in water, $\mu_p'$, in units of the nuclear magneton, $\mu_n$ or $\mu_N$,
since all the other ingredients in the calculation were known much more accurately. 

The experiments that were taken into consideration in 1969 were as follows:
\begin{align}
\begin{array}{llll}
\mbox{Date} & \mu'_p/\mu_n & R_E & \mbox{Refs.}\cr\hline
\mbox{1949--51} & 2.792690(30) & 1836.096(20) &\mbox{\cite{exp1}}\cr
\mbox{1950--6} & 2.79267(10) & 1836.083(66) &\mbox{\cite{exp2}}\cr
\mbox{1955--63}& 2.792701(73) &1836.104(48) & \mbox{\cite{exp3}}\cr
\mbox{1961}& 2.792832(55) &1836.190(36) & \mbox{\cite{exp4}}\cr
1965& 2.792794(17) & 1836.165(11) & \mbox{\cite{exp5}}\cr
1967& 2.792746(52) & 1836.133(34)& \mbox{\cite{exp6}}\cr
1967&2.79260(13)& 1836.037(85)&\mbox{\cite{exp7}} \cr
\hline
\end{array}
\end{align}
These data show a significant tension between the two most accurate measurements, in 1949--51 and 1965, of perhaps
$3\sigma$. Several pages of \cite{1969} were devoted to a detailed analysis of this tension, which resulted in a final
recommendation much closer to the earlier result than the later, of $1836.109(11)$. It can be seen that the proposed
mixing of $R_I$ and $R_G$ with $\sin^2\lambda\approx .22$ resolves this tension in a different way.

The main reason for the adjustment in 1973 was the publication in 1971 of a comprehensive catalogue
of accurate atomic masses \cite{LGSmith} measured in 1965. This catalogue effectively calibrated
inertial and gravitational masses against each other, to significantly greater accuracy than had
ever been done before.
The other experiments that were taken into consideration in 1973 were:
\begin{align}
\begin{array}{llll}
\mbox{Date} & \mu'_p/\mu_n & R_E & \mbox{Refs.}\cr\hline
\mbox{1949--51} & 2.792690(30) & 1836.096(20) &\mbox{\cite{exp1}}\cr
1965& 2.792794(17) & 1836.165(11) & \mbox{\cite{exp5}}\cr
1970 & 2.792783(16) & 1836.158(11)& \mbox{\cite{exp8}}\cr
1972 & 2.792786(17) & 1836.160(11)& \mbox{\cite{exp9}}\cr
1972 & 2.792777(20) & 1836.154(13)& \mbox{\cite{exp10}}\cr
1972 & 2.7927738(12) & 1836.152(01)& \mbox{\cite{exp11}}\cr
1972 & 2.7927748(23) & 1836.152(01)& \mbox{\cite{exp12}}\cr
\hline
\end{array}
\end{align}
In this case, the decision was taken to ignore all the experiments except the two most accurate, both dating from 1972, both
giving the currently accepted value of $R_I$ to an accuracy better than 1ppm.

Clearly, the evidence presented here is insufficient to claim definitive proof of the failure of the equivalence principle
at the elementary particle level. But the evidence is consistent with the model that I propose, which suggests that either
some new experiments, or a different interpretation of experiments from the 1970s to 1990s, might be useful. In particular,
the fact that electro-weak mixing in the standard model can apparently separate the inertial mass ratio from the
gravitational mass ratio surely indicates that the weak force is closely related to gravity.
Since weak interactions always emit neutrinos and/or anti-neutrinos, it is plausible that these particles have a role to play in
the propagation of the gravitational field, as suggested elsewhere in this paper, and in other proposed discrete models
\cite{finite}. 

\subsection{A possible explanation}
The values of $R_E$ that are reported depend not only on the experiment, but also on the theory that is used for 
interpreting the experiment. In \cite{1969} analysis was done both classically (labelled WQED---without quantum electrodynamics)
and with QED. After \cite{1973} the full standard model was used, since this was necessary
in order to justify the increasingly precise values that were being obtained from experiment. 
 
Classical Newtonian mass is defined by $\rep 4a$ alone, and calibrated by weighing objects in
the Earth's gravitational field. 
It is therefore reasonable to call this gravitational mass.
The discovery of electricity, and the development of the theory of electromagnetism in the
19th century, led to an electromagnetic field in $\rep 3a+\rep 3b$, in such a way that the combination of charge
and mass appears in $\rep3a\otimes \rep 3b=\rep 4a+\rep 5$. At this stage a calibration of $\rep 5$
against $\rep 4a$ is implicitly required in order to maintain a single concept of mass, rather than separating out
two components. 
Let us therefore label the $\rep 5$-component `inertial mass', while being careful
to understand that the term `inertial mass' as used in other contexts may well be a mixture of the two components
distinguished here.

The classical (WQED) analysis therefore uses a mass value for each particle that is a compromise
between a theoretical gravitational mass in $\rep 4a$ and a theoretical inertial mass in $\rep5$.
On the other hand, the 
QED analysis uses the Dirac spinor in $\rep4b$ to produce a mass value in the square of the spinor
 \begin{align}
 \rep 4b\otimes \rep 4b=\rep 1+\rep 3a+\rep 3b+\rep 4a+\rep 5.
 \end{align}
  By projecting out the unwanted terms,
 this representation can be identified with the classical one, so that the two can be calibrated against each other \cite{1969,1973}.
  
The full standard model introduces $\rep 2a+\rep 2b$ also into the Dirac spinor, so that the gravitational mass can be
recovered in $\rep 2a\otimes \rep 2b=\rep 4a$. This provides enough information to separate the two different types of mass
in $\rep 4a$ and $\rep 5$. The standard model can therefore work entirely with the $\rep 5$-mass, defined by a further
projection from QED, and now defined as inertial mass, or simply mass.
This pushes the calibration of the (gravitational) $\rep 4a$-mass against the (inertial) 
$\rep5$-mass into the electro-weak mixing angle that defines the
relationship between the spinors of types $\rep2a + \rep 2b$ and $\rep 4b$. 
From this point onwards, therefore, the standard model can, and does, ignore gravity completely.
Nevertheless, the icosahedral model shows how it may be possible to recover a gravitational mass, distinct from
inertial mass, from the
unified electro-weak theory, in such a way that the weak interaction becomes one ingredient in 
a quantum theory of gravity.

\subsection{Parameters of the Lorentz gauge}
The discrete model offers two distinct real versions of the Dirac algebra, corresponding to a single complex version
in the standard model. Therefore the $16$ different real $2$-spaces each have their own complex structure, defined by
some mixing angle between the real and complex parts. Six of these effectively identify the two symmetry groups,
that is $Spin(3,1)$ acting on $\rep 4b$ and $Spin(4)$ acting on $\rep 2a+\rep 2b$,
while the other $10$ are parameters that must appear somewhere else in the standard model. 

These ten parameters can be mapped to two scalars and two vectors
under the action of $SO(3,1)$ on adjoint $SL(2,\HH)$, 
so that the two vectors can be identified with the Cabibbo--Kobayashi--Maskawa (CKM) matrix 
\cite{Cabibbo,KM}
and the Pontecorvo--Maki--Nakagawa--Sakato (PMNS) matrix
\cite{Pontecorvo,MNS}. Each matrix 
holds four real parameters of $\rep 4a$ in a $3\times 3$ matrix representing $\rep 3a\otimes \rep 3b$.
The two scalars probably represent the electroweak mixing angle
and the fine structure constant, although the former can also be interpreted as a mass ratio,
so might appear elsewhere, in which case we might replace it with the strong
coupling constant. Of course, all these parameters depend on a particular choice of identification of
$SO(3,1)$ with $SO(4)$, which can only be constant on a subgroup $SO(3)$, so that they all vary with the energy scale,
as experiment confirms.

In addition, the complex scalar in the Dirac algebra is used for the mass term, so that each of the other $15$ dimensions of
the algebra gets its own 
mass value from this mixing. While these naturally split into $3+3+4+5$, the symmetry-breaking splits
$4$ as $1+3$ and $5$ as $1+1+3$ complex dimensions, so that they can be identified with $3+3+3+3$ masses for the
elementary fermions, and $1+1+1$ for the $Z$, $W$ and Higgs bosons, to give the $15$ fundamental masses of
the standard model. Here we 
see again that the masses of the elementary particles might be a compromise between inertial and gravitational masses,
and hence gauge-dependent.
It is known experimentally that the $Z/W$ mass ratio varies according to the energy. 
For quarks in general, and for neutrinos and the Higgs boson,
the masses are
so difficult to measure accurately that there is no real evidence either way. That leaves just three charged lepton masses,
for which the experimental evidence is that they are genuinely constant. This is consistent with a 
choice of constant $SO(3)$.

\section{Speculations}
\subsection{The Koide formula}
\label{Koide}
The empirical Koide formula \cite{Koide} relates the masses of the three generations of electron as follows:
\begin{align}
\frac{m_e+m_\mu+m_\tau}{(\sqrt{m_e}+\sqrt{m_\mu}+\sqrt{m_\tau})^2} &\approx \frac23.
\end{align}
Although there is no generally accepted reason why this formula should hold, it has been predictive of more accurate
values of the $\tau$ mass,
and it has not been experimentally falsified yet.
Writing $x,y,z$ for the square roots of the masses, the formula
can be re-arranged into the form
\begin{align}
x^2+y^2+z^2&\approx 4(xy+xz+yz).
\end{align}

In the group algebra model, masses for all three generations lie in the representation $\rep5$. Moreover, the generations are labelled
by the representation $\rep3b$. Since $S^2(\rep3b)=\rep1+\rep5$, we might expect to find the square roots of the masses in 
$\rep3b$.
If we therefore take coordinates $x,y,z$ for $\rep3b$, with the permutation $(x,y,z)$ acting as a generation symmetry for electrons,
then we have coordinates 
$x^2,y^2,z^2,xy,xz,yz$
 for $\rep1+\rep5$, in which $x^2+y^2+z^2$ spans the scalar.
The element of order $3$ fixes also the vector $xy+xz+yz$. Hence there is some multiple of $xy+xz+yz$ that is equal to the
sum of the masses of the three generations, that is $x^2+y^2+z^2$. 

Why this multiple should be $4$, I do not know, but I suspect it
comes somehow from the relationship between $\rep5$ and $\rep1+\rep4a$.
On the other hand, it may come from the relationship between the permutation
representation $\rep1+\rep5$ and either $\rep3a+\rep3b$ or $\rep6$, both of which are monomial representations on $6$ points.

Now there is an experimental problem, that while both the Koide formula and my formula (relating the three electron masses to the 
proton and neutron masses) are consistent with experiment, they are not consistent with each other 
(as pointed out to me by Marni Sheppeard in April 2015). They both offer 
predictions of the mass of the tau particle to at least $8$ significant figures, of which only the first $4$ agree.
In terms of the model, the representation $\rep5$ contains not only the three generations of electron, but two more particles that I have
identified as the proton and the neutron. By reducing the representation modulo $3$ (see Section~\ref{mod3}), we can separate the
mass from the charge into $\rep1+\rep4a$ and obtain my formula. 
But to get the Koide formula, we have to split the representation 
into leptons and baryons, as $3+2$. 

There is no mathematical operation that achieves this, other than
breaking the symmetry by restricting to a subgroup. This has the effect of ignoring the difference in mass between the
proton and the neutron. Hence the Koide formula is only approximately correct. The fact that it is such an extraordinarily good
approximation is entirely due to the fact that the proton and neutron masses are so nearly equal. If the above conjecture
regarding the origin of this mass difference is accepted, then we could say that the (historical and philosophical, not
physical) reason for its accuracy is that there
are so many days in the year!

\subsection{Variable gravity}
The change that 
I have proposed making to the Riemann Curvature Tensor effectively mixes the three generations of electron, and therefore
has the effect that the ratio of gravitational to inertial mass for the electron is not constant, but depends on the local values of the tensor.
If we take the inertial mass to be fixed, as it is in the standard model, then this means that the gravitational mass of an electron
can vary according to the local conditions. This obviously has a very small effect on the actual gravitational force on ordinary matter,
since there is such a small range of conceivable values for the electron mass. Nevertheless, these effects may be large enough to be
detected by the right experiments.

Even if we consider the whole range from $0$ up to the difference between the neutron and proton masses, the effect could never be
bigger than $.1\%$. The actual range that might be achievable in experiments may be considerably smaller than this. There are in fact
at least four independent experimental anomalies that might in principle be explained an effect of this kind. These are
\begin{itemize}
\item inconsistent measurements of Newton's gravitational constant $G$;
\item galaxy rotation curves inconsistent with Newtonian gravity;
\item the flyby anomaly;
\item the Pioneer anomaly.
\end{itemize}
Many proposals have been made to explain these anomalies, and some are regarded as adequately explained already.
But if the current proposal can explain them all simultaneously, then it may be considered more satisfactory 
than having separate \emph{ad hoc} explanations
for each one.

Consider first the measurements of $G$
\cite{Cavendish,Gillies,newG,QLi}. Here we might expect to find different values of $G$ for different materials, based
essentially on the proportion of the total mass that can be attributed to electrons. For typical heavy metals this proportion
does not vary much from one element to another, but for lighter elements, including iron, it is significantly larger. Thus we might
expect slightly different results in the two cases, since both lighter and heavier elements have
been used in actual experiments. The extremal difference amounts to roughly $3.6$ electrons per iron atom, which is 
approximately 40ppm. 

Now if we look at the whole range of potential gravitational masses, this translates to a difference
anywhere from $0$ to 100ppm. If both masses in the experiment are affected, this could double to 200ppm.
This range is sufficiently large to cover the reported anomalies, and suggests that it is worth taking
a more detailed look at the individual
experiments, to see if there are indeed correlations of the kind I suggest between the materials used in the experiment
and the values of $G$ that are reported in the results. If so, one can then try to see if the magnitudes of the anomalies
are consistent with an explanation of the kind proposed. 

It may indeed be possible to interpret such a difference (if it is confirmed experimentally) as a magnetic effect of gravity.
That is, the proposed new quantum process can happen in the nucleus of a metal atom as the result of an interaction
with a very low energy neutrino, which is absorbed and re-emitted without any change in energy, but with a minute change
in the direction of the momentum. If this is a valid interpretation, then it may also allow us to interpret an atom undergoing
$\beta$ decay, with the emission of an anti-neutrino, as a magnetic monopole. Indeed, one might therefore regard
the neutron itself as a magnetic monopole, and $\beta$ decay the evidence for it.

Next consider galaxy rotation curves. Here the standard LCDM ($\Lambda$ cold dark matter) model invokes `dark matter'
as a hypothetical constituent of galaxies, but no such matter has ever been detected in the laboratory. Alternative theories
such as MOND (modified Newtonian dynamics) are empirical laws
\cite{Milgrom1,Milgrom2,Milgrom3}, without any physical explanation, but reportedly
do a better job of describing what is actually observed \cite{MOND,MOND2,Banik}. 
There are various different ways of looking at MOND-type models,
many of which are equivalent to a failure of the equivalence principle in some form, either as a variable gravitational-to-inertial
mass ratio, or a variable gravitational `constant' $G$. My model can be interpreted in either of these ways, but is actually
more subtle than either of them. Since in this model the electron gravitational mass (but not its inertial mass)
depends on an angle between two different rotations, 
it is the rotation of the galaxy itself that is key to the understanding of the effects normally attributed to dark matter.

Now it is possible to use these ideas to estimate the critical acceleration below which the MOND effects become
noticeable. For this, we need to appreciate that our own rotation within the Milky Way affects our own understanding
of the local gravitational fields that we measure. Since this rotation is effectively ignored in our modelling of gravity
within the Solar System, our models of gravity cannot be more accurate than this. In other words, below a critical
acceleration roughly equal to our own acceleration towards the centre of the Milky Way, both Newtonian gravity and
general relativity break down. This is indeed what is observed. 

If this is really true, it
means that the entire basis of our theories of gravity is misconceived, and that mass is
an \emph{effect} of gravity, not the \emph{cause}.  It is \emph{rotation} that is the cause of gravity, not the other way around. 
This provides a whole new paradigm for gravity \cite{newparadigm}, for which we seek new experimental evidence.
Now the flyby anomaly \cite{flyby,Hafele}
arises when
a spacecraft flies close to the Earth in order to pick up speed from the motion of the Earth around the Sun.
In a number of instances, the increase in speed is observed to be greater than that predicted by the theory.
But in some instances, no anomalous increase in speed was detected, and in one case the anomaly had the opposite sign.

An empirical law describing the effect was obtained, that was dependent on the effective rotation of the spacecraft
in the North-South direction. Now rotations are described by the representation $\rep3a$ or $\rep3b$, so that the two rotations
of the Earth and the spacecraft are perpendicular vectors in (say) $\rep3a$, and the gravitational effect of these two
rotations lies in $S^2(\rep3a)=\rep1+\rep5$. The magnitude of the effect  
lies in the spin $2$ representation,
and is proportional to
\begin{align}
\cos I - \cos O &= 2\sin((O+I)/2)\sin((O-I)/2),
\end{align}
where $I$ and $O$ are the effective latitudes of the inbound and outbound trajectories.
The fact that this formula is a product of sines is what puts it into $S^2(\rep3a)$.

Finally, let us consider the Pioneer anomaly \cite{Pioneer1,Pioneer2}. This was an effect that was detected in the two Pioneer
spacecraft, once they had finished their manoeuvres in the outer Solar System and were left to coast on out of the
Solar System altogether. An anomalous acceleration towards the Sun, of the same magnitude as the critical acceleration
mentioned above, was detected. While it is nowadays generally accepted that the anomaly can be explained by 
greater experimental uncertainties and systematic biases than were originally built into the analysis, there is also the
possibility that the effect was real, and that the experiment actually did detect the rotation of the Solar System around the
centre of the galaxy.

No discussion of variable gravity can be considered complete without mention of dark energy or the cosmological constant.
Recent experiments \cite{Hubble1,Hubble2,nodarkenergy} do indeed cast doubt on the existence of this concept,
although it is an essential part of standard cosmology. In the group algebra model, two scalars in $\rep1+\rep5$ are replaced
by vectors in $\rep3a+\rep3b$. One of them represents dark matter, and has been conclusively dealt with in the
preceding discussion. The other one represents dark energy, which likewise does not exist in the proposed model.
It is replaced, again, by another rotation, presumably
on an even larger scale. 

\subsection{Muons}
\label{muong-2}
The model under discussion here is relevant to the calculation of $g-2$ for the muon. There are two competing calculations,
one \cite{muontheory,muonHVP} using ordinary quantum chromodynamics (QCD), the other \cite{Fodor} using lattice QCD.
The former is in tension with experiment \cite{muong-2}, while the latter 
is consistent with experiment.

Now the proposed model essentially contains lattice QCD, after breaking the symmetry down to $SL(2,3)$ so that the
cubic lattice in $3$-space is preserved. This has the effect that $\rep3a$ and $\rep3b$ become identical, so that a complex structure
can be put onto $\rep3a+\rep3b$, and a symmetry group $SU(3)$ imposed on it. 
However, the continuous group $SU(3)$ is not a subgroup of the group algebra,
and does not preserve the essential quantum structure of the elementary particles. Hence, using this 
(unphysical) symmetry group
is liable to produce the wrong answer.

Therefore my model is consistent with both lattice QCD and experiment, while continuous QCD disagrees with all three
independent methodologies. As a result it would be reasonable to conclude that QCD is definitively falsified. The error can
moreover be located to the imposition of an inconsistent complex structure onto $\rep3a+\rep3b$. The calculations then proceed in
what should be $\rep3a\otimes \rep3b=\rep4a+\rep5$, but after symmetry breaking is $3\otimes 3 = 1+3+2+3$. This is the same
symmetry-breaking that pertains in lattice QCD, but in continuous QCD, the representation $3+2+3$ forms 
an irreducible adjoint representation of $SU(3)$. These symmetries, however, are inconsistent with the $\rep4a+\rep5$ structure,
and are therefore unphysical.

The magnitude of the error can be estimated from the fact that the distinction between $\rep3a$ and $\rep3b$ is also of
crucial importance for gravity, as described in the previous section. Since the muons in the experiment are rotating in a $15$-meter
diameter ring, and since rotation causes gravity, as we have seen above, there is a mixing of gravity with QCD in this
experiment. The effect is much the same as in the flyby anomaly, and in the CP-violation of neutral kaon decays,
so that there is a factor of $\sin\theta$, where $\theta\approx 2.36\times 10^{-6}$ radians. In other words, the ratio
between the two competing values of muon $g-2$ is approximately $.99999764$.

\section{Conclusion}
In this paper I have examined the hypothesis that there are close connections between the representation theory of the
binary icosahedral group, on one hand, and various models that attempt to go beyond the standard model of particle physics,
on the other. Many such
connections have been found, but the question remains: is there a deeper meaning to these connections, or not? 
One or two general
principles have emerged from this investigation, the first of which is that, as a general rule, 
symmetric tensors describe matter and structures,
while anti-symmetric tensors describe forces. 

The second is a correspondence between permutation representations,
in particular $\rep1+\rep4a$ and $\rep1+\rep5$, and monomial representations, here $\rep5$ 
and $\rep3a+\rep3b$ respectively. The monomial representation $\rep5$
introduces a triplet symmetry to the fundamental concept of mass, and hence provides a place to model the three generations.
The monomial representation $\rep3a+\rep3b$ introduces a doublet symmetry to the fundamental particles, extending their number
from $6$ to $12$.  The forces then do not depend on whether we use the permutation representation or the monomial representation, since
\begin{align}
\Lambda^2(\rep1+\rep4a)&=\rep3a+\rep3b+\rep4a\cr
&=\Lambda^2(\rep5)\cr
\Lambda^2(\rep1+\rep5)&=\rep3a+\rep3b+\rep4a+\rep5\cr
&=\Lambda^2(\rep3a+\rep3b).
\end{align}
Matter in the two cases appears to be described by
\begin{align}
S^2(\rep4a)&=\rep1+\rep4a+\rep5\cr
S^2(\rep5)&=\rep1+\rep4a+\rep5+\rep5.
\end{align}
It then appears that there is not as much difference as might have been thought, 
between the $\rep4a$ case, on which the theory of relativity can be built, and the $\rep5$ case, 
on which the standard model of particle physics can be built.

A third principle is the reduction of representations modulo $2$ (which again relates $\rep1+\rep5$ to $\rep3a+\rep3b$),
modulo $3$ (which relates $\rep1+\rep4a$ to $\rep5$) and modulo $5$ (which relates $\rep1+\rep3a$ and $\rep1+\rep3b$ to $\rep4a$).
One of the main themes of this paper has been the similarities and differences between $\Lambda^2(\rep1+\rep5)$ and $S^2(\rep5)$. 
For many purposes they seem to be interchangeable, but they are different representations and therefore have
different physical interpretations. I have suggested the former for boson labels, and the latter for fermion labels.
Supersymmetry can perhaps be thought of as an apparent (but mathematically inconsistent, and therefore not real)  
correspondence between
the two.

General relativity is described by the field strength tensor in $\Lambda^2(\rep4a)$ and
the stress-energy tensor in $S^2(\rep4a)$, but appears to lack the force term in $\rep4a$, so that gravity is interpreted not as a force,
but as curvature in spacetime (itself described by the $\rep4a$ representation), via the Riemann Curvature Tensor in
$S^2(\Lambda^2(\rep4a))$.
This model therefore suggests that it may be possible to restore gravity to the status of a force, by extending $\Lambda^2(\rep4a)$
to $\Lambda^2(\rep1+\rep4a)=\Lambda^2(\rep5)$. Furthermore, by extending again to $\Lambda^2(\rep1+\rep5)$, we obtain the same force
tensor as in particle physics. 
In other words, the incompatibility of general relativity with particle physics seems to have
disappeared, at least at the level of the finite symmetries. 

Of course, this is not yet a unified theory, and there is no guarantee that the Lie groups can be made to match up as
closely as the finite group representations. But it does suggest a new place to try to build the foundations for 
such a theory. Moreover, even in its current very rudimentary form, the model makes predictions about kaon decays
that are different from the standard model predictions, some of which are known to contradict experiment.
Therefore the model offers a revised description of kaons that can be tested experimentally against the standard model
description. Finally, I would like to point out that my model correctly postdicts \emph{six} apparently fundamental dimensionless
`constants', all to an accuracy of $.05\%$ or better, by relating them to a gravitational gauge group.

As far as quantum gravity and cosmology are concerned, the model suggests that a modification of general relativity is
required in order to incorporate three generations of electrons into the description of the matter in the universe, and the
effective gravitational force that arises as a consequence. I have shown that dark matter can be effectively modelled
as the \emph{differences} between the three generations, in such a way that it does not consist of particles, and can never
be detected as particles. I have shown, moreover, the importance of taking into account rotations in the modelling of gravity,
and therefore of mass, as well as the other fundamental forces \cite{Speake}. 
Such rotations appear to explain many things that cannot be explained in any other way.

\end{document}